\documentclass[aps,prx,reprint,groupedaddress,longbibliography]{revtex4-2}

\usepackage{graphicx}
\usepackage{amsmath,amssymb,amsfonts,bm}
\usepackage{mathtools}
\usepackage{siunitx}
\usepackage[usenames,dvipsnames]{color}

\newcommand{\bra}[1]{\left\langle #1 \right|}
\newcommand{\ket}[1]{\left| #1 \right\rangle}
\newcommand{\bracket}[2]{\langle #1 | #2 \rangle}
\newcommand{\tr}{\operatorname{tr}}
\renewcommand{\Re}{\operatorname{Re}}
\renewcommand{\Im}{\operatorname{Im}}

\def\bk{\bm{k}}
\def\bbeta{\bm{\beta}}
\def\nn{\nonumber}

\begin{document}

\title{ Amoeba Formulation of Non-Bloch Band Theory in Arbitrary Dimensions }


\author{Hong-Yi Wang}
\affiliation{Institute for Advanced Study, Tsinghua University, Beijing 100084, China}

\author{Fei Song}
\affiliation{Institute for Advanced Study, Tsinghua University, Beijing 100084, China}

\author{Zhong Wang}
\email{wangzhongemail@tsinghua.edu.cn}
\affiliation{Institute for Advanced Study, Tsinghua University, Beijing 100084, China}


\date{\today}

\begin{abstract}

The non-Hermitian skin effect dramatically reshapes the energy bands of non-Hermitian systems, meaning that the usual Bloch band theory is fundamentally inadequate as their characterization. The non-Bloch band theory, in which the concept of Brillouin zone is generalized, has been widely applied to investigate non-Hermitian systems in one spatial dimension. However, its generalization to higher dimensions has been challenging. Here, we develop a formulation of the non-Hermitian skin effect and non-Bloch band theory in arbitrary spatial dimensions, which is based on a natural geometrical object known as the amoeba. Our theory provides a general framework for studying non-Hermitian bands beyond one dimension. Key quantities of non-Hermitian bands, including the energy spectrum, eigenstates profiles, and the generalized Brillouin zone, can be efficiently obtained from this approach.

\end{abstract}

\maketitle

\section{\label{sec:intro}Introduction}

Non-Hermitian Hamiltonians have wide applications in many branches of physics, ranging from classical wave phenomena to open quantum systems \cite{Ashida2021,Bergholtz2021RMP}. Among various non-Hermitian systems, those that have spatially periodic structures are especially important and extensively studied. Recently, their interplay with topological physics has stimulated the fruitful investigations of non-Hermitian topological states \cite{Bergholtz2021RMP, Ding2022}. One of the key phenomena uncovered in this direction is the non-Hermitian skin effect (NHSE), which refers to the counterintuitive feature that the nominal ``bulk eigenstates'' of a non-Hermitian Hamiltonian are exponentially localized at the boundary \cite{yao2018edge, kunst2018biorthogonal, lee2018anatomy, alvarez2017, Helbig2019NHSE, Xiao2019NHSE, Ghatak2019NHSE, Wang2022morphing}. Unlike the topological boundary modes whose number scales linearly with the boundary area, the number of skin modes is proportional to the system volume. Among other consequences, the NHSE implies that the energy spectra of a non-Hermitian system can be drastically different under the experimentally favored open boundary condition (OBC) and the theoretically efficient periodic boundary condition (PBC). In sharp contrast to the Anderson localization, the NHSE-induced exponential localization can occur in pristine non-Hermitian systems without any disorder.

As a pronounced deviation from the usual Bloch-wave picture, the NHSE implies that the standard Bloch band theory is insufficient to characterize a generic non-Hermitian band. For example, the OBC energy spectrum cannot be calculated within this framework. To address this serious issue, a non-Hermitian generalization of the standard band theory, known as the non-Bloch band theory, has been introduced and applied to various non-Hermitian systems \cite{yao2018edge,Yokomizo2019,Longhi2019Probing,Longhi2020chiral, Kawabata2020nonBloch,Yang2019Auxiliary,Song2019real,Lee2020Unraveling,Yi2020}. A central concept in this theory is the generalized Brillouin zone (GBZ), which is the proper surrogate for the conventional Brillouin zone (BZ) in Hermitian bands. The GBZ allows efficient computation of the continuous OBC energy spectrum of a large system without the need of diagonalizing a large Hamiltonian matrix in real space. Meanwhile, the shape of the GBZ directly tells the information of real-space eigenstates, e.g., the skin localization lengths. Moreover, the topological invariants on the GBZ are able to predict the number of topological boundary modes of non-Hermitian systems, whereas the usual BZ-based topological numbers fail to do so.

So far, the GBZ has been well defined only in one dimension (1D). Finding a general definition and formula for GBZ in two and higher dimensions is challenging, because the well-known 1D approach is not amenable to a straightforward generalization \cite{yao2018edge,Yokomizo2019,sczhangmemorial}. In certain special cases, the two-dimensional (2D) GBZ has been approximately defined and calculated \cite{yao2018chern,liu2019second}. However, a general approach without resorting to uncontrolled approximations has been lacking.

In this paper, we present a general formulation of non-Hermitian band theory in arbitrary dimensions. Among other results, it tells how the GBZ and related spectral properties are quantitatively determined beyond 1D. In 1D, our formulation reduces to the well-known GBZ formulation. The formulation is universal in the sense that our main results are applicable to models in arbitrary spatial dimensions, and with arbitrary degrees of freedom in a unit cell.

Our theory is based on a natural geometrical object called the amoeba by mathematicians \cite{Gelfand1994,viro2002amoeba,Forsberg2000laurent,theobald2002computing,rullgaard2001polynomial}.  Inspired by the amoeba and related mathematical tools, we formulate a theory which characterizes the NHSE quantitatively in arbitrary spatial dimensions. In particular, it is possible to directly calculate the energy spectrum and density of states (DOS) in the thermodynamic (i.e., large-size) limit without the troublesome finite-size errors. We show in a theorem that the energy spectrum can be obtained from the shape of the amoeba. 
We also demonstrate, despite the geometry-dependent NHSE, the existence of a universal spectrum (amoebic spectrum) to which the OBC spectrum under any generic geometry converges. Furthermore, the amoeba inspires a definition and the associated algorithm of the GBZ in arbitrary spatial dimensions. Among other applications, this amoeba-based GBZ provides a starting point for calculating non-Bloch band topology beyond 1D.

The remainder of this article is arranged as follows. In Sec.~\ref{sec:inspir}, we go through the existing method of determining the GBZ in 1D, and then try to find clues for its higher-dimensional generalization.  In Sec.~\ref{sec:math} we introduce the basic mathematical properties of the amoeba and the associated Ronkin function, which will be useful in calculating the DOS and the GBZ. In Secs.~\ref{sec:dos}-\ref{sec:gbz}, we introduce the amoeba formulation for non-Hermitian systems, and then make use of this formulation and the theory of Toeplitz matrices to establish a universal way to determine the DOS as well as the GBZ.   Non-Bloch band topology based on the proposed GBZ theory is studied in Sec.~\ref{sec:topo}. Finally, in Sec.~\ref{sec:ineq}, several useful inequalities on the OBC and PBC spectra are proved from the amoeba approach.

\section{\label{sec:inspir}Motivation}

\subsection{Review of 1D non-Bloch band theory}

We start with reviewing the concept of GBZ in 1D \cite{yao2018edge,Yokomizo2019,Yang2020Auxiliary,Zhang2020correspondence}, searching for clues of its higher-dimensional generalizations.

A general 1D tight-binding Hamiltonian with OBC can be written as
\begin{equation}
    H = \sum_{i,j=1}^{L}\sum_{a,b} \ket{i,a} \left( t_{j-i} \right)_{ab} \bra{j,b}   ,
\end{equation}
where $i,j$ are the position indices, and $a,b$ are the indices for intracell degrees of freedom (band indices). Explicitly, the matrix $H$ looks like
\begin{equation}
    H = 
    \begin{pmatrix}
        t_0    & t_1 & t_2 &     & \cdots & & 0  \\
        t_{-1} & t_0 & t_1 & t_2 &  & &    \\
        t_{-2} & t_{-1} & t_0 & t_1 & \ddots & & \vdots\\
               & t_{-2} & t_{-1} & t_0 & \ddots & & \\
        \vdots &  & \ddots & \ddots & \ddots & & t_{2} \\
        &&&&& t_0 & t_{1}   \\
        0   &   & \cdots & & t_{-2} & t_{-1} & t_0    \\
    \end{pmatrix},
\end{equation}
in which each $t_n$ stands for a square matrix $\left( t_{n} \right)_{ab}$. As the hopping matrix $t_{j-i}$ depends only on the spatial distance $j-i$, we have translational symmetry in the bulk. The Hermiticity condition $t_n=t_{-n}^\dagger$ is not required since non-Hermitian Hamiltonians are our focus. A finite hopping range $n_c$ is assumed so that the hopping matrix $t_{n}=0$ when $|n| >n_c$.  Given this real-space Hamiltonian, the corresponding Bloch Hamiltonian is the Fourier transform of $t_n$:
\begin{equation}
    h(e^{ik}) = \sum_n t_n \left( e^{ik} \right)^n.
\end{equation}
Note that the Bloch Hamiltonian has been written as $h(e^{ik})$ instead of the more frequently used $h(k)$. This simplifies our notations when the real-valued wave vector $k$ is generalized to the complex plane, which amounts to making the substitution $e^{ik}\rightarrow e^{\mu+ik}=\beta$ ($k$ and $\mu$ are real-valued). Evidently, $h(\beta)=\sum_n t_n\beta^n$ is a matrix-valued Laurent polynomial of $\beta$.

To solve the real-space Schrödinger equation $H\ket{\psi} = E\ket{\psi}$, we first note that the equations have identical form at all spatial coordinates $j$ (except some points near the boundary):
\begin{equation}    \label{bulk_schrodinger}
    E \psi_j = \sum_{n} t_n \psi_{j+n}  .
\end{equation}
Thanks to the translational invariance in the bulk, the Schrödinger equation is a linear recurrence equation with constant coefficients. The standard approach is to take a trial solution $\psi_j = v \beta^j$ ($v$ is a vector whose dimension is the number of bands), and the bulk equation gives
\begin{align}
    E v \beta^j = \sum_{n} t_n v \beta^{j+n}   ,    \\
    \Longleftrightarrow \, [E-h(\beta)] v = 0   .
\end{align}
If this trial solution can appear as a component of an eigenstate wave function, it is necessary that 
\begin{equation}    \label{eq:char}
    \det[E-h(\beta)] = 0   ,
\end{equation}
which is called the characteristic equation for this problem. The most general form of a wave function satisfying the bulk equations is then a linear superposition of functions $v_n (\beta_n)^{j}$, where $\beta_n$ is a solution of $\det [E-h(\beta)] = 0$ and $v_n$ is the corresponding eigenvector. Thus, an eigenstate of energy $E$ is expressed as
\begin{equation}    \label{1d_general_sol}
    \ket{\psi} = \sum_{j=1}^{L} \sum_a \sum_{n=1}^{M+N} c_{n} (v_n)_a (\beta_n)^j \ket{j,a},   
\end{equation}  
where the coefficients $c_n$ are to be determined by the boundary conditions. We expand the characteristic polynomial as $\det [E-h(\beta)] = a_{-M}(E) \beta^{-M}+\dots+a_N(E) \beta^N$ with $a_{-M} (E)$, $a_N (E)$ nonzero, and sort its roots as $|\beta_1(E)| \leq |\beta_2(E)| \leq \dots \leq |\beta_{M+N}(E)|$. It has been found that, in the thermodynamic (large-size) limit,  the OBC boundary condition results in the following simple equation, known as the GBZ equation \cite{yao2018edge,Yokomizo2019,Zhang2020correspondence}:
\begin{equation}    \label{eq:1dequal}
    \left| \beta_{M}(E) \right| = \left| \beta_{M+1}(E) \right|.
\end{equation}
It turns out that all the $\beta_M, \beta_{M+1}$ solutions form a closed curve called the GBZ on the complex plane of $\beta$, which contains key information about the eigenstate profiles, including the conventional wave vector $k$ and the spatial decay rate $\mu$ of a skin mode. Once the GBZ is obtained, one can insert $\beta\in \mathrm{GBZ}$ into Eq.~\eqref{eq:char} to obtain the OBC energy spectrum. Furthermore, the topological boundary modes are dictated by the topological invariants defined in the GBZ rather than in the BZ. This phenomenon is known as non-Bloch bulk-boundary correspondence \cite{yao2018edge,Yokomizo2019}. Thus, in many senses, GBZ plays a similar role as BZ does in Hermitian systems. Band theory based on the GBZ concept is known as the non-Bloch band theory. 

The condition $|\beta_M(E)| = |\beta_{M+1}(E)|$ lies at the heart of the 1D non-Bloch band theory. It can be intuitively justified with the simplest example, in which there are only one band and two $\beta$'s, $\beta_1$ and $\beta_2$, with $|\beta_1|\leq |\beta_2|$. Then, Eq. (\ref{1d_general_sol}) reads $\ket{\psi} = \sum_{j=1}^L\psi_j \ket{j}$ with $\psi_j=c_1  \beta_1^j + c_2  \beta_2^j$. The open boundary condition amounts to adding fictitious sites $j=0$ and $j=L+1$ at the left and right ends, with $\psi_0=0$ and $\psi_{L+1}=0$, respectively. At the left end, we find $\psi_0=0 \Longrightarrow c_1 = -c_2$. Then, at the right end, it is possible for the two terms to cancel each other only if $|\beta_1| = |\beta_2|$; otherwise, the $\beta_2^{L+1}$ term would be much larger than the $\beta_1^{L+1}$ term. In the general case, a rigorous treatment is to write down the system of linear equations for the boundary conditions, and it turns out that the coefficient matrix is an $(M+N)\times(M+N)$ square matrix. Letting its determinant vanish results in $|\beta_M(E)| = |\beta_{M+1}(E)|$ \cite{Yokomizo2019,Zhang2020correspondence}. 

Unfortunately, we will see that in higher dimensions, the above treatment relying on the small rank of the coefficient matrix will become quite intractable. This has been a major obstacle in the attempt to study GBZ in higher dimensions.

\subsection{The way to the amoeba}

\begin{figure}[t]
    \centering
    \includegraphics[width=\linewidth]{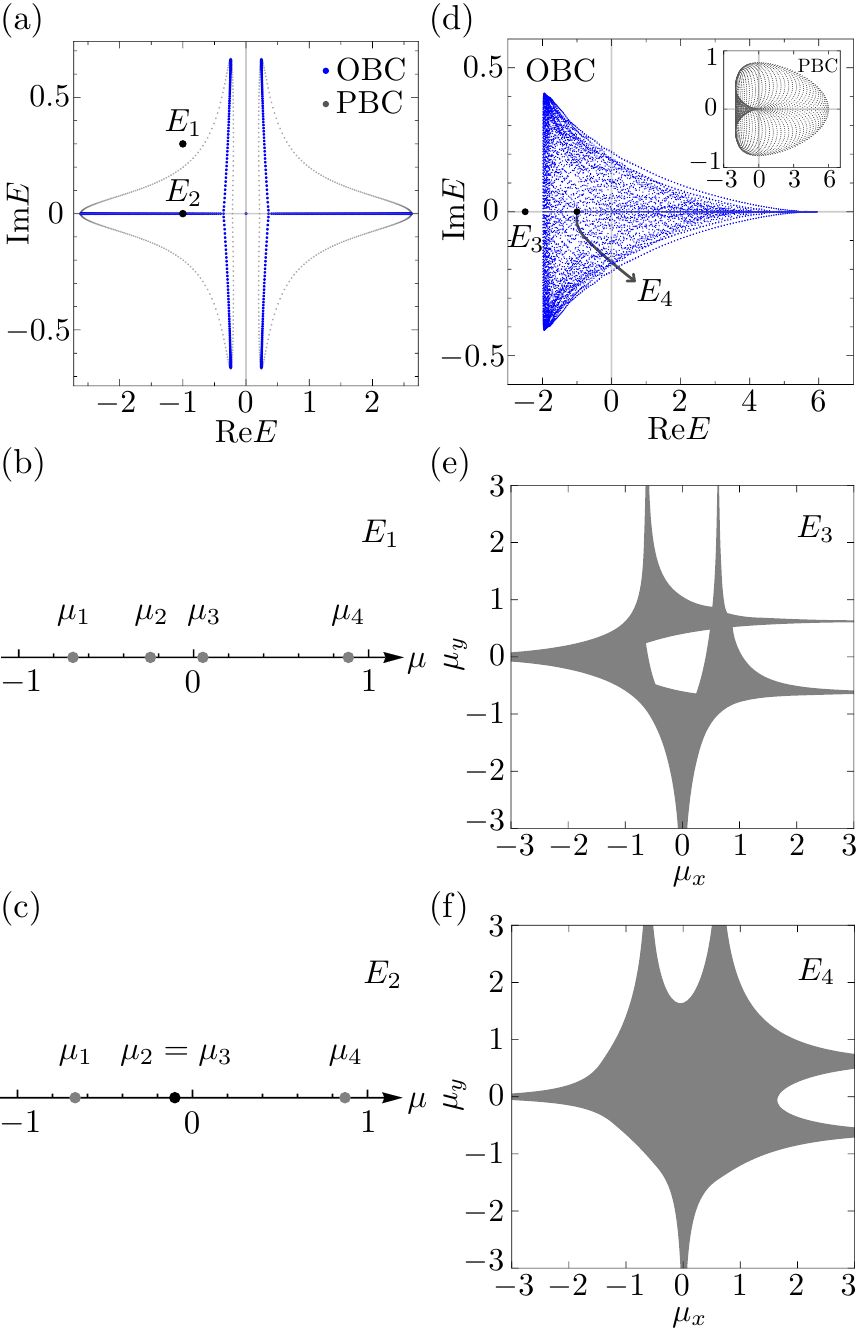}
    \caption{Energy spectra and amoebae. (a) Energy spectrum of the 1D non-Hermitian SSH model Eq.~(\ref{eq:SSH}) for a chain with length $L=300$, under OBC (blue) and PBC (gray), respectively. Parameter values are $t_1=t_2=1$, $t_3=0.7$, and $\gamma=4/3$. (b)(c) Illustrations of $\mu=\log|\beta|$, where $\beta$ satisfies the 1D characteristic equation $\det [E-h(\beta)]=0$. $E$ is taken to be $E_1=-1+0.3i$ in (b), and $E_2=-1$ in (c). For $E_2$ belonging to the OBC spectrum, we have $\mu_2(E_2)=\mu_3(E_2)$.  (d) Energy spectrum of the 2D model Eq.~\eqref{eq:dgmodel}. Main figure: disk geometry (OBC) with diameter $L=140$. Inset: torus geometry (PBC). Parameter values are $t=1$, $t'=0.5$, and $\gamma=0.2$. (e)(f) Illustrations of the 2D amoebae whose points are $(\mu_x,\mu_y)=(\log|\beta_x|,\log|\beta_y|)$, where $\bm{\beta}$ satisfies the 2D characteristic equation $\det [E-h(\bm{\beta})]=0$ of the model Eq. (\ref{eq:dgmodel}). $E_3=-2.5$ in (e), and $E_4=-1$ in (f). Notably, there is a hole in the amoeba for $E_3$ outside the OBC spectrum, and no hole for $E_4$ in the OBC spectrum. }
    \label{fig:amoeba}
\end{figure}

To demonstrate the applications of non-Bloch band theory, we consider the non-Hermitian Su-Schrieffer-Heeger (SSH) model with the Bloch Hamiltonian \cite{yao2018edge,kunst2018biorthogonal}:
\begin{align}
    h(e^{ik})  &= \left[ t_1+ ( t_2+t_3 ) \cos k \right] \sigma_x \nn \\
   &\phantom{=} + \left[ (t_2-t_3) \sin k +i\dfrac{\gamma}{2} \right] \sigma_y,
\end{align}
or, in terms of $\beta$,
\begin{align}    \label{eq:SSH}
    h(\beta) &= \left[ t_1+ \frac{t_2+t_3}{2} (\beta+\frac{1}{\beta}) \right] \sigma_x \nn\\
    &\phantom{=}  + \left[\frac{t_2-t_3}{2i}(\beta -\frac{1}{\beta}) +i\dfrac{\gamma}{2} \right] \sigma_y,
\end{align}
where $\sigma_{x,y,z}$ are the Pauli matrices. It is known that the eigenstates exhibit NHSE under OBC, and consequently, the OBC and PBC energy spectra are drastically different \cite{yao2018edge, kunst2018biorthogonal}. As an illustration, the OBC and PBC spectra are plotted in Fig.~\ref{fig:amoeba}(a), for parameter values $t_1=t_2=1$, $t_3=0.7$, and $\gamma=4/3$. This choice of parameters is in the topologically nontrivial regime, and therefore topological edge modes with $E=0$ are found in the OBC energy spectrum. The characteristic equation Eq.~\eqref{eq:char} is a quartic equation of $\beta$, and its four solutions are shown in Figs.~\ref{fig:amoeba}(b) and (c), for $E=E_1=-1+0.3i$ and $E=E_2=-1$, respectively. Instead of $\beta$ itself, we show $\mu=\log|\beta|$, namely the imaginary part of the wave vector. As stated in Eq.~(\ref{eq:1dequal}), when and only when $\left| \beta_{2}(E) \right| = \left| \beta_{3} (E)\right|$, i.e., $\mu_2(E)=\mu_3(E)$, will $E$ belong to the OBC energy spectrum. This is the case for $E=E_2$ [Fig.~\ref{fig:amoeba}(c)]. The corresponding $\beta_2(E_2)$ and $\beta_3(E_2)$ belong to the GBZ.

Remarkably, the GBZ equation, Eq. (\ref{eq:1dequal}), enables us to find the OBC energy spectra and other physical quantities without the need of diagonalizing a large real-space Hamiltonian whose size grows with the system size. The thermodynamic-limit quantities are obtained directly from the GBZ equation. A natural question arises: What is the higher-dimensional counterpart of Eq. (\ref{eq:1dequal})?

\begin{figure}[t]
    \centering
    \includegraphics[width=\linewidth]{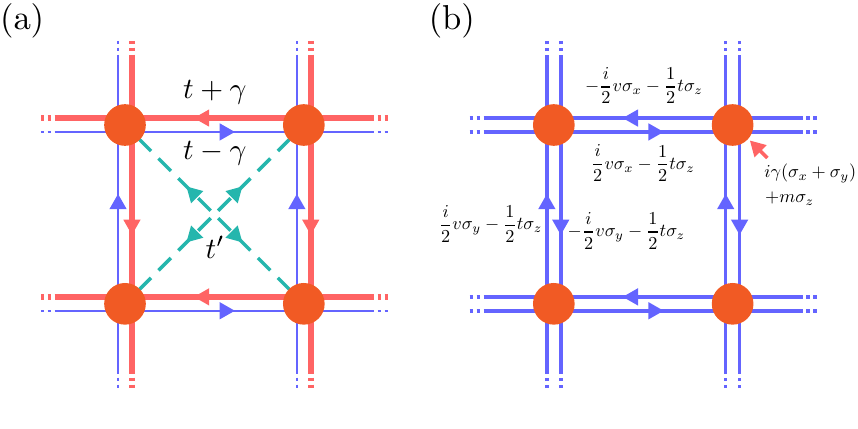}
    \caption{Illustration of the real-space hopping of the 2D models used in this article. (a) Single band model Eq.~(\ref{eq:dgmodel}). (b) The non-Hermitian Chern-band model Eq.~(\ref{eq:qwz}). All the parameters $t,\gamma,v,m$ are real-valued. }
    \label{fig:hopping}
\end{figure}

To be concrete, we consider a single-band model shown in Fig.~\ref{fig:hopping}(a). With the notation $\bm{\beta}=(\beta_x,\beta_y)$, the corresponding Bloch Hamiltonian is
\begin{align}   \label{eq:dgmodel}
    h(\bm{\beta}) &= t \left( \beta_x+\beta_x^{-1}+\beta_y+\beta_y^{-1} \right) \notag\\
    &\phantom{=} + t' \left( \beta_x+\beta_x^{-1} \right) \left( \beta_y+\beta_y^{-1} \right) \notag\\
    &\phantom{=} + \gamma \left( \beta_x-\beta_x^{-1}+\beta_y-\beta_y^{-1} \right),
\end{align} in which the $\gamma$ terms generate nonreciprocal hoppings [see Fig.~\ref{fig:hopping}(a)].
For concreteness, we fix parameters $t=1$, $t'=0.5$, and $\gamma=0.2$.  The energy spectra from brute-force numerical diagonalization are shown in Fig.~\ref{fig:amoeba}(d).

We hope to generalize the non-Bloch band theory to 2D, such that the energy spectrum can be obtained from the GBZ instead of the real space. The characteristic equation takes the same form as in 1D (and the reasoning is also the same):
\begin{equation}
    \det [E-h(\bm{\beta})] = 0.
\end{equation}
For our single-band model, the determinant can be dropped, and therefore the characteristic equation is simply $E-h(\bm{\beta})=0$. Notably, the zero locus of $\det[E-h(\bm{\beta})]$, namely the $\bm{\beta}$-solution space of $\det[E-h(\bm{\beta})]=0$, is (real) two dimensional. In fact, the solution space can be locally parameterized by the complex-valued $\beta_x$ or $\beta_y$, one of which is determined by the other via the characteristic equation. In contrast, the zero locus of $E-h(\beta)$ for a 1D system is zero dimensional, that is, several isolated $\beta$ points.

The plausible next step, following the approach in the 1D non-Bloch band theory, is to select the legitimate $\beta$'s by adding the boundary conditions. For 1D systems with OBC, the number of constraint equations imposed by the boundary condition does not grow with the system size $L$, which greatly eases the derivation of the GBZ equation, Eq.~\eqref{eq:1dequal} \cite{yao2018edge,Yokomizo2019}. For 2D OBC systems (e.g., with square or disk geometry), however, the number of constraint equations is proportional to the linear size $L$. It is therefore challenging to exploit all these boundary-condition equations. Consequently, it is difficult to obtain a 2D counterpart of Eq.~(\ref{eq:1dequal}) from the approach similar to 1D. 

Although a straightforward generalization to 2D looks infeasible, we can still find some clues from the 1D GBZ construction. Equation (\ref{eq:1dequal}) and  Figs.~\ref{fig:amoeba}(b) and \ref{fig:amoeba}(c) suggest that the moduli of the solutions to the characteristic equation contain much useful information about the energy spectrum. Correspondingly, we plot the solutions to the 2D characteristic equation, followed by mapping $(\mu_x,\mu_y)=(\log|\beta_x|,\log|\beta_y|)$, in Figs.~\ref{fig:amoeba}(e) and \ref{fig:amoeba}(f). This geometrical object is known as the amoeba in mathematics literature (see Sec.~\ref{sec:math} for an introduction) \cite{Gelfand1994,viro2002amoeba,Forsberg2000laurent,theobald2002computing,rullgaard2001polynomial}. We notice that a hole exists in the amoeba in Fig.~\ref{fig:amoeba}(e), for which the energy $E_3$ does not belong to the OBC spectrum. In contrast, there is no hole in the amoeba in Fig.~\ref{fig:amoeba}(f), with energy $E_4$ belonging to the OBC spectrum. Viewed from this amoeba perspective, 1D non-Hermitian energy spectra also exhibit similar behaviors. In 1D, the amoeba consists of discrete points, and the hole is simply the open interval between two adjacent points. For example, the open interval $(\mu_2,\mu_3)$ in Fig.~\ref{fig:amoeba}(b) can be viewed as a hole, which closes in Fig.~\ref{fig:amoeba}(c) with $\mu_2=\mu_3$.

From the above examples, we observe that the absence (presence) of a hole in the amoeba of the characteristic polynomial could be an indicator of the energy $E$ being (not being) in the OBC energy spectrum. This is a key observation of the present work. To obtain more quantitative results from this observation, it is helpful to know some mathematical properties about the amoeba.

\section{\label{sec:math}Mathematical properties of the amoeba and Ronkin function}

In this section, we shall introduce the basic concept of amoeba and a closely related analytic tool, the Ronkin function. As a quite recent concept in mathematics, the amoeba was introduced by Gelfand \textit{et al.} in 1994 \cite{Gelfand1994}. Albeit elementary, the notion of amoeba has deep connections with various concepts in algebraic geometry, which has stimulated extensive studies in mathematics \cite{viro2002amoeba,Forsberg2000laurent,theobald2002computing,rullgaard2001polynomial}.

Let $f$ be a Laurent polynomial of $\beta_j$, $j=1,2,\dots,d$, where $d$ will be identified as the spatial dimension in our study. The amoeba of $f$ is defined as the log-moduli of the zero locus of $f$,
\begin{equation}
    \mathcal{A}_f = \left\lbrace \log\left|\bm{\beta}\right|: f(\bm{\beta})=0 \right\rbrace \subset \mathbb{R}^d,
\end{equation}
in which we use the notation $\log|\bm{\beta}| \coloneqq (\log|\beta_1|, \dots, \log|\beta_d|)$ to simplify our expressions. Similar notations such as $e^{\bm{\mu}} \coloneqq (e^{\mu_1}, \dots, e^{\mu_d})$ are used hereafter. In our case, the Laurent polynomial in use is $\det[E-h(\bm{\beta})]$. We can see that the geometric objects in Figs.~\ref{fig:amoeba}(b)(c) and (e)(f) are 1D and 2D amoebae, respectively.

The name amoeba was motivated by its appearance in 2D: It has slim ``tentacles'' extending to infinity, and sometimes several ``vacuoles'' (holes) inside its body. Importantly, a particular hole plays an important role in our formulation. It is known that the amoeba in any spatial dimensions is a closed set, and each hole is a convex set \cite{Forsberg2000laurent}.

A useful analytic tool in the study of amoeba is the Ronkin function, which is defined as \cite{viro2002amoeba,Passare2004,ronkin1974introduction}
\begin{equation}    \label{Rf}
    R_f(\bm{\mu}) = \int_{T^d} \left( \dfrac{d\theta}{2\pi} \right)^d
    \log\left| f (e^{\bm{\mu}+i\bm{\theta}}) \right|,
\end{equation}
where the domain of integration is the $d$-dimensional torus $T^d=[0,2\pi]^d$, and the expression is simplified by the notations $f(e^{\bm{\mu}+i\bm{\theta}}) \coloneqq f(e^{\mu_1+i\theta_1},\dots, e^{\mu_d+i\theta_d})$, and $ (d\theta/2\pi)^d \coloneqq (d\theta_1/2\pi) \dots (d\theta_1/2\pi)$.

It is beneficial to study the gradient of the Ronkin function \cite{Forsberg2000laurent,Passare2004}. To this end, we can express the integrand in $R_f$ as $\log|f|=\Re \log f$. The real part can be taken at the end of the calculation. It turns out that the integral is real-valued before taking the real part, and therefore the ``$\Re$'' symbol can be discarded. The derivation proceeds as
\begin{align}
    \nu_j
    &= \frac{\partial R_f (\bm{\mu})}{\partial \mu_j}  \nn\\
    &= \Re \int_{T^d} \left( \dfrac{d\theta}{2\pi} \right)^d
    \partial_{\mu_j} \log  f (e^{\bm{\mu}+i\bm{\theta}}) \nn \\
    &= \Re \int_{T^d} \left( \dfrac{d\theta}{2\pi} \right)^d
    \frac{ \partial_{\mu_j} f(e^{\bm{\mu}+i\bm{\theta}} )}{f(e^{\bm{\mu}+i\bm{\theta}})} \nn\\
    &= \Re \int_{T^d} \left( \dfrac{d\theta}{2\pi} \right)^d
    \frac{ -i \partial_{\theta_j} f(e^{\bm{\mu}+i\bm{\theta}} )}{f(e^{\bm{\mu}+i\bm{\theta}})}.  \label{gradient}
\end{align}
In the last line, we use $\partial_{\mu_j} = -i\partial_{\theta_j}$ acting on $f$. We observe that
\begin{equation}    \label{winding}
    w_j
    =  \frac{1}{2\pi i} \int_0^{2\pi} d\theta_j  \frac{\partial_{\theta_j} f(e^{\bm{\mu}+i\bm{\theta}}) }{f(e^{\bm{\mu}+i\bm{\theta}})}
\end{equation}
is the winding number of the phase of $f$ along a circle parametrized by $\theta_j$, and therefore it is always real-valued. Thus, the gradient $\nu_j$ is the average of the winding number $w_j$ on the $(d-1)$-dimensional torus parametrized by $(\theta_1,\dots,\theta_{j-1},\theta_{j+1},\dots,\theta_d)$:
\begin{equation}
    \nu_j
    =  \int_{T^{d-1}}  \frac{d\theta_1\dots d \theta_{j-1}d\theta_{j+1}\dots d\theta_d}{(2\pi)^{d-1}} w_j .
\end{equation}
For example, in 2D, one has $\nu_1=\int_0^{2\pi}\frac{d\theta_2}{2\pi}  w_1$ and $\nu_2=\int_0^{2\pi}\frac{d\theta_1}{2\pi} w_2$.

The next fact we see is that the Ronkin function is exactly linear on each connected component of the complement of the amoeba. We call each of these components (either bounded or unbounded) a hole of the amoeba. In fact, when $\bm{\mu}$ is not in the amoeba, $f(e^{\bm{\mu}+i\bm{\theta}})\neq 0$ is satisfied in the entire $T^d$ parametrized by $(\theta_1,\dots,\theta_d)$. It follows that $w_j$ is a constant integer as $\{\theta_1,\dots,\theta_{j-1},\theta_{j+1},\dots,\theta_d\}$ vary, and therefore the average $\nu_j$ is the same integer. Thus, we can assign to each amoeba hole an integer tuple $\bm{\nu} = (\nu_1, \dots, \nu_d)$ dubbed the order of the hole. The orders of two different holes cannot be the same \cite{Forsberg2000laurent}. Crucially, there exists at most one hole with order $\bm{\nu}=(0,\dots,0)$, which we call the central hole. Because the order is zero, the Ronkin function is a constant in this hole.  Moreover, the Ronkin function is convex in the entire $\bm{\mu}$ space; i.e., $R_f(\lambda\bm{\mu}_1 + (1-\lambda)\bm{\mu}_2)\leq  \lambda R_f(\bm{\mu}_1) +  (1-\lambda) R_f(\bm{\mu}_2)$ is satisfied for any two points $\bm{\mu}_{1,2}$ and $0<\lambda<1$ \cite{viro2002amoeba,Passare2004}. An easy corollary of the convexity is that a Ronkin function converges everywhere: If it were $-\infty$ at one point, convexity would imply that it is $-\infty$ everywhere.

\begin{figure}[t]
    \centering
    \includegraphics[width=\linewidth]{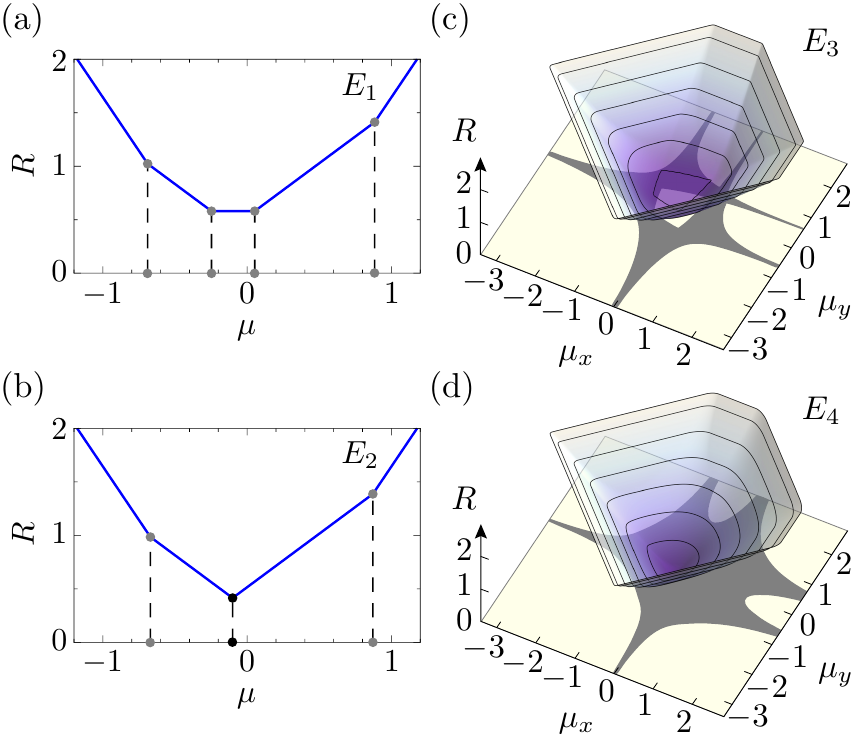}
    \caption{The Ronkin function in (a, b) 1D and (c, d) 2D, taken at energies (a) $E_1$, (b) $E_2$, (c) $E_3$, and (d) $E_4$ stated in Fig.~\ref{fig:amoeba}. The Bloch Hamiltonian is Eq.~(\ref{eq:SSH}) for (a, b); it is Eq.~(\ref{eq:dgmodel}) for (c, d). The parameter values are the same as stated in Fig.~\ref{fig:amoeba}. In both 1D and 2D, the Ronkin function is strictly linear on each component (hole) of the complement of the amoeba, where the gradient equals the integer index. The Ronkin function is always convex. Consequently, when the central hole exists, the minimum is reached on the central hole; otherwise, the minimum is reached at a single point in the amoeba. }
    \label{fig:ronkin}
\end{figure}

In 1D, the Ronkin function is closely related to Jensen's formula in complex analysis \cite{stein2010complex}, which reads
\begin{equation}
    \dfrac{1}{2\pi} \int_0^{2\pi} d\theta \, \log\left| g\left( R e^{i\theta} \right) \right|
    = \log\left| g(0) \right| + \sum_{k=1}^l \log\left| \dfrac{R}{z_k} \right|, \label{Jensen}
\end{equation}
where $g$ is a holomorphic function with $g(0)\neq 0$, and $z_k$ ($k=1,\dots,l$) are the zeros of $g$ enclosed by the circle $|\beta|=R$.  Jensen's formula can be readily obtained from Eq. (\ref{gradient}) with $d=1$, in which case the gradient $\nu=w$ (the index $j=1$ is redundant). In fact, the left-hand side of Eq. (\ref{Jensen}) is exactly the Ronkin function $R_g(\log|R|)$. To calculate it, we order the zeros of $g$ as $|z_1|\leq|z_2|\leq|z_3|\leq\cdots$. According to Eqs.~(\ref{gradient}) and (\ref{winding}), the gradient $\nu=\frac{\partial R_g}{\partial (\log|R|)}$ equals the winding number of $g$ along the circle $|\beta|=R$, which counts the number of enclosed zeros. Therefore, we have $\nu=0$ for $\log|R|<\log|z_1|$, and $\nu=k$ for $\log|z_{k}|<\log|R|<\log|z_{k+1}|$. It follows that, when the circle $|\beta|=R$ encloses $l$ zeros, 
\begin{equation}    \label{ronkinjensen}
    \begin{aligned}
        R_g(\log|R|) &= \log|g(0)| + \log \frac{|z_2|}{|z_1|} + 2\log\frac{|z_3|}{|z_2|}+3\log\frac{|z_4|}{|z_3|}   \\
        &\phantom{=} +\dots  + l\log \frac{|R|}{|z_l|}     \\
        &= \log\left| g(0) \right| + \sum_{k=1}^l \log\left| \dfrac{R}{z_k} \right|,
    \end{aligned}
\end{equation} 
which is exactly Jensen's formula Eq. (\ref{Jensen}).

We now apply the explicit formula of the Ronkin function to $\det[E-h(\beta)] = a_{-M}(E) \beta^{-M}+\dots+a_N(E)\beta^N$, whose roots are $|\beta_1(E)| \leq |\beta_2(E)| \leq \dots \leq |\beta_{M+N}(E)|$. We rewrite it as $\det[E-h(\beta)]=f(\beta)g(\beta)$ with $f(\beta)=a_{-M}(E)\beta^{-M}$, so that $g(\beta)$ has no pole in the complex plane and $g(0)=1$. Applying Eq. \eqref{ronkinjensen} to $g$, we have
\begin{align}   \label{eq:rk_explicit_1d}
    R_{\det(E-h)} (\mu)
    &= R_{f}(\mu) + R_g (\mu) \nn \\   &= \log |a_{-M}| - M\mu + \sum_{k=1}^{l} ( \mu - \log | \beta_k |  ), \nn \\
\end{align}
for $\log|\beta_l| \leq \mu \leq \log|\beta_{l+1}|$. Thus, $R_{\det(E-h)}$ is a piecewise linear function of $\mu$ in 1D. To relate to the concept of amoeba, we note that each $\log|\beta_k|$ is a component of the amoeba, and each interval $(|\beta_k|,|\beta_{k+1}|)$ is an amoeba hole. Particularly, the open interval $(|\beta_M|,|\beta_{M+1}|)$ is exactly the central hole on which the Ronkin function is flat, because taking a derivative on Eq.~\eqref{eq:rk_explicit_1d} we find $\partial R_{\det(E-h)} / \partial\mu = -M+M = 0$. 
Therefore, Eq. \eqref{eq:1dequal} means that an energy $E$ belongs to the OBC bulk spectrum when the central hole shrinks to zero size. Thus, the Ronkin function provides crucial information about the energy spectra.  Figures \ref{fig:ronkin}(a) and \ref{fig:ronkin}(b) are two examples of the 1D Ronkin function. The Ronkin function is flat in the central hole of Fig.~\ref{fig:ronkin} (a) corresponding to $E$ outside the energy spectrum, and the hole shrinks to a point in Fig.~\ref{fig:ronkin} (b) corresponding to $E$ in the energy spectrum.

In 2D and higher dimensions, it is challenging to obtain a closed form for the Ronkin function. Nevertheless, regardless of the spatial dimensions, the Ronkin function is always globally convex, and is linear in the amoeba holes, i.e., connected components of the complement of the amoeba. If there is a central hole, the Ronkin function takes minimum in the entire hole, namely, the function has a flat bottom in this hole [Fig.~\ref{fig:ronkin} (c)], otherwise the Ronkin minimum is reached at a single point in the amoeba [Fig.~\ref{fig:ronkin} (d)].

In view of the aforementioned relation between the GBZ and Ronkin function in 1D, it is natural to ask whether there is a deep connection between the non-Hermitian energy spectra and the Ronkin function in higher dimensions. In fact, we have already seen some numerical clues for such a connection. Our observation about 2D non-Hermitian systems at the end of Sec.~\ref{sec:inspir} can be rephrased in terms of the Ronkin function: If and only if the minimum of the Ronkin function $R_{\det(E-h)}$ is reached at a single point $\bm{\mu}_{\min}(E)$, instead of in a hole with nonzero size, will $E$ be in the OBC bulk spectrum.

In 1D, the single point where the Ronkin function takes the minimum is $\mu_{\min}(E)=\log|\beta_M(E)|=\log|\beta_{M+1}(E)|$. Note that $|\beta_M(E)|=|\beta_{M+1}(E)|$ is the decay factor of an OBC eigenstate.  Thus, the location of the minimum of the Ronkin function precisely determines the decay factor of an OBC eigenstate. In other words, the Ronkin function tells the shape of the GBZ. We propose that this reformulation of GBZ in terms of the Ronkin function is generalizable to non-Hermitian systems in higher dimensions, which is justified in the following sections.

\section{\label{sec:dos}Energy Spectra and density of states}

\begin{figure*}[t]
    \centering
    \includegraphics[width=431pt]{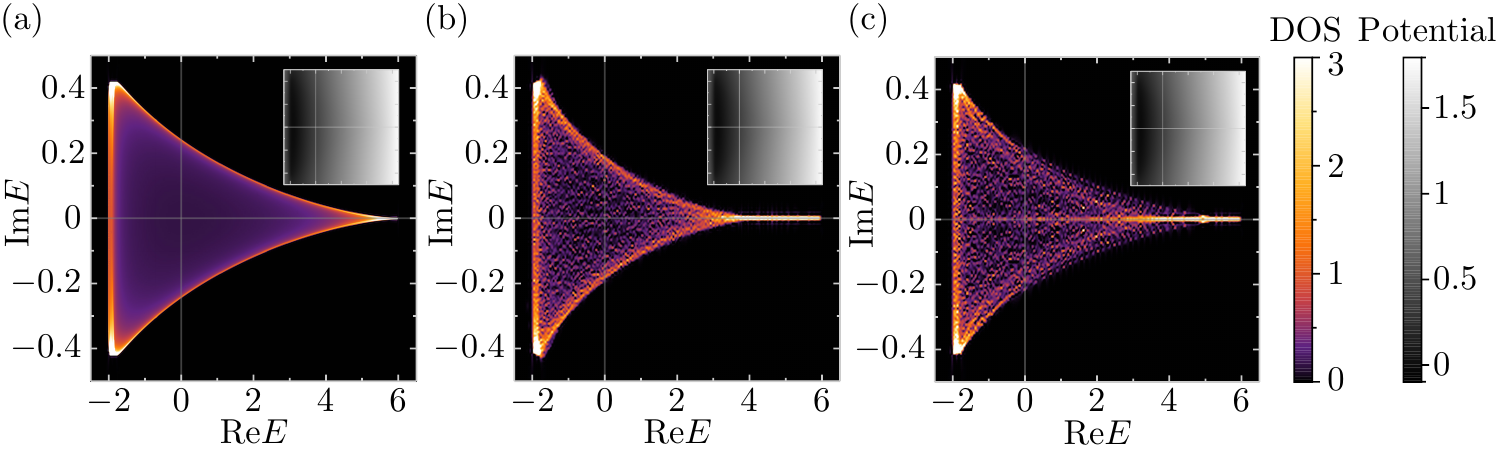}
    \caption{DOS from the Ronkin function and diagonalization of the real-space Hamiltonian. The Hamiltonian used here is Eq. (\ref{eq:dgmodel}) [Fig.~\ref{fig:hopping}(a)], with parameter values $t=1$, $t'=0.5$, and $\gamma=0.2$. (a) DOS from the Ronkin function, via Eq. (\ref{eq:dos}).  (b) DOS from diagonalizing the real-space Hamiltonian on a square with side length $L=130$. An on-site random potential distributed uniformly in $[-0.5,0.5]$ is added at each boundary site. (c) DOS from diagonalizing the real-space Hamiltonian on a disk with diameter $L=140$ (without random potential at the boundary).   The insets show the Coulomb potential, $\phi(E)$ in (a),  and $\Phi(E)$ in (b) and (c). To facilitate comparison with (a), the DOS in (b) and (c) is also obtained from the Coulomb potential via Eq. (\ref{eq:dos-Phi}), in which $\Phi(E)$ is generated by diagonalizing the real-space Hamiltonian.  }
    \label{fig:dos}
\end{figure*}

We now introduce the amoeba formulation of non-Hermitian energy band theory in $d$ spatial dimensions. One of our main objectives is to calculate the DOS associated with the energy spectrum, which is defined as the number of states per area on the complex energy plane, divided by the volume of the system, in the thermodynamic limit. Because the DOS has an $O(L^d)$ volume denominator, only the bulk states are relevant in the thermodynamic limit. All contributions from edge states, bound states, etc., vanish in the thermodynamic limit. For example, the number of possible surface states grows with the size as $O(L^{d-1})$, which contributes $O(1/L)$ that vanishes in the thermodynamic limit. Thus, we focus on the bulk spectrum.

\subsection{Statement of the proposal}

To describe the DOS of complex energy, it is convenient to use the language of electrostatics. Let us assign electric charge $-1/N$ to each energy eigenvalue $\epsilon_n$, where $N\sim L^d$ is the total number of unit cells (in this convention the total charge sums up to the number of energy bands, which is independent of the size $N$). In terms of Dirac's $\delta$ function, the DOS of complex energy can be written as $\rho(E)= \lim_{N\rightarrow\infty} \sum_n\frac{1}{N}\delta(E-\epsilon_n)$, which is just the absolute value of electric charge density.

Given the electric charge, the corresponding Coulomb potential $\Phi(E)$ is given by 
\begin{align}
    \Phi(E)
    &= \dfrac{1}{N} \sum_{\epsilon_n}\log\left| E-\epsilon_n \right| \notag\\
    &\xrightarrow[N\to\infty]{} \int d^2 E' \, \log\left|E-E'\right| \rho(E'). \label{coulomb}
\end{align}
Conversely, the DOS, or the absolute value of the charge density in the electrostatics language, can be readily obtained from the Coulomb potential by taking the Laplacian on the complex energy plane:
\begin{equation}    \label{eq:dos-Phi}
    \rho (E) = \dfrac{1}{2\pi}\Delta\Phi (E),
\end{equation} where $\Delta=\frac{\partial^2}{\partial (\Re E)^2} + \frac{\partial^2}{\partial (\Im E)^2}$, and the $N\to\infty$ limit is taken for $\Phi(E)$.

One of our main proposals is that the Coulomb potential $\Phi (E)$, in the $N\rightarrow\infty$ limit, can be obtained from the Ronkin function,
\begin{equation}
    \Phi(E)=\phi(E), \label{Phiphi}
\end{equation}
where $\phi(E)$ is the minimum of the Ronkin function of $\det(E-h)$ by varying $ \bm{\mu}$:
\begin{equation} \label{phiE}
\phi(E) = \min_{\bm{\mu}} R_{\det(E-h)} \left( \bm{\mu} \right),
\end{equation} or $\phi(E) =   R_{\det(E-h)} \left( \bm{\mu}_\text{min} \right)$, where $\bm{\mu}_\text{min}$ minimizes the function.
Therefore, the DOS can be directly obtained from the Ronkin function,
\begin{equation}    \label{eq:dos}
    \rho (E) = \dfrac{1}{2\pi}\Delta\phi (E).
\end{equation}

In the special cases that the spectrum is a 1D object consisting of lines or curves in the complex energy plane, it is more preferable to define the DOS as states per length rather than states per area. In fact, the states per area will diverge in these cases, while states per length is in general finite. This is the case for 1D non-Hermitian systems, and higher-dimensional systems with real spectra. In a small neighborhood of a segment of the 1D spectrum, suppose that $\bm{n}$ is a normal vector to the segment. From the electrostatic analogy and Eq.~(\ref{eq:dos}), it is easy to see that the DOS per length denoted by $\rho_{\mathrm{1D}}$ is given by
\begin{equation} \label{1D}
    \rho_{\mathrm{1D}} = \dfrac{1}{2\pi} \left| \left( \dfrac{\partial \phi}{\partial \bm{n}} \right)_+ - \left( \dfrac{\partial \phi}{\partial \bm{n}} \right)_- \right| ,
\end{equation}
where the two derivatives are taken on opposite sides of the curve segment.

\subsection{Numerical evidence}

Before diving into a thorough analytic approach, we provide numerical evidence for the Ronkin-function-based formula, Eq.~(\ref{eq:dos}). We take the model Eq. (\ref{eq:dgmodel}) [Fig.~\ref{fig:hopping}(a)] as an example. In Fig.~\ref{fig:dos}, we numerically compare the DOS derived from the Ronkin function, via Eq.~(\ref{eq:dos}), and that from diagonalizing the OBC Hamiltonian. The Laplacian is implemented by discretizing the complex energy plane into grid points. The DOS obtained from the Ronkin function agrees well with that from real-space Hamiltonians, though there are some differences that can be naturally attributed to the finite-size nature of the real-space calculations. First, the spectra from real-space diagonalization in Figs.~\ref{fig:dos}(b) and (c) are slightly narrower in the $\Im E$ direction than the one derived from the Ronkin function in Fig.~\ref{fig:dos}(a). Second, the eigenvalues from real-space diagonalization seem more likely to concentrate on the real axis. These differences can be explained by the non-Bloch \textit{PT} symmetry, which features a unique size dependence of non-Hermitian spectra in two and higher dimensions \cite{song2022nonbloch}. In fact, our real-space Hamiltonian [see Eq. (\ref{eq:dgmodel}) and Fig.~\ref{fig:hopping}(a)] has real-valued hoppings and therefore commutes with the complex conjugation operator: $[H,\mathcal{K}]=0$. Combined with the NHSE, it means that the model has non-Bloch \textit{PT} symmetry, which implies real energy spectra when the size is small. As the size $L$ grows to infinity, the proportion of real eigenenergies diminishes to zero \cite{song2022nonbloch}.  Since the length $L$ adopted in Fig.~\ref{fig:dos}(b) and (c) is finite, the \textit{PT} symmetry breaking is incomplete, leaving a nonzero proportion of real eigenenergies.

We emphasize that the Ronkin function tells the unique universal spectrum and universal DOS of the OBC system, the precise meaning of which we explain below. It is evident that the amoeba and the Ronkin function do not use information about the shape (e.g., square or disk) or boundary details (e.g., clean or locally perturbed) of the OBC system. The Ronkin function yields the same DOS regardless of geometrical details. Therefore, this approach intrinsically assumes a detail-independent (therefore, universal) spectrum with a universal DOS, which is supported by our numerical calculation. Indeed, our numerical results indicate that the DOS of an OBC system with an arbitrary shape converges in the large-size limit to the same universal DOS, at least when certain random local perturbations are added to the boundary [e.g., in Fig.~\ref{fig:dos}(b)]. Note that the DOS thus obtained is independent of the specific forms of boundary randomness \footnote{We have also checked that adding local randomness in the bulk has a similar effect. Here, the bulk randomness should be sufficiently weak such that the resulting energy spectrum reflects the properties of the pristine Hamiltonian. Specifically, one can add random potential only to $N_d$ unit cells, with $N_d/N\to 0$ (though $N_d\rightarrow\infty$) when taking the large-size limit $N\to\infty$ ($N$ is the number of unit cells). }. In generic cases, even the boundary randomness is not necessary to ensure the convergence to the universal DOS. For example, the DOS of disk geometry without boundary random potential already resembles the universal DOS [Fig.~\ref{fig:dos}(c)]. In contrast, for a polygon geometry (e.g., a square), boundary randomness significantly helps the DOS to converge to the universal DOS. Without the boundary randomness, the polygon geometry can exhibit the geometry-dependent non-Hermitian skin effect, by which different polygons may have different DOS \cite{zhang2022universal}. Our intuitive understanding of this phenomenon is as follows. The boundary of a polygon consists of straight line segments, which are perfectly reflective in the sense that the wave vector (momentum) parallel to the line segment is conserved during wave reflection. Thus, the boundary fails to fully mix waves with different wave vectors, and therefore the Hamiltonian can be viewed as fine-tuned rather than generic. As such, the spectrum exhibits the fingerprint of specific geometry rather than the universal spectral properties of the Hamiltonian. Boundary or bulk randomness breaks the wave vector conservation and couples waves with different wave vectors, which generates a more generic energy spectrum. An analogous phenomenon is the critical non-Hermitian skin effect in 1D \cite{Li2020critical,Yokomizo2021scaling}. In the zero-coupling limit of two coupled 1D chains, the straightforward application of the GBZ equation, Eq. (\ref{eq:1dequal}), does not yield the correct OBC energy spectrum \footnote{When taking the zero-coupling limit, one sets the chain length to be large and fixed.}. The zero-coupling limit represents a fine-tuned point, and a small interchain coupling brings the spectrum to that predicted by the GBZ theory \cite{Li2020critical,Yokomizo2021scaling}. In 2D, our numerical results suggest that the Hamiltonian should be viewed as fine-tuned for polygon shapes, and a small randomness restores the universal spectrum characterized by the universal DOS.

To summarize, there exists a geometry-independent universal spectrum that can be calculated from the amoeba and Ronkin function. By nature, it can be called the ``amoebic spectrum.''  The DOS of an OBC system with a generic shape always approaches the universal DOS in the large-size limit. When the DOS of an OBC system with a certain (nongeneric) shape appears to deviate from the universal DOS, this deviation can be eliminated by adding a small random local perturbation. From an experimental point of view, the universal spectrum is particularly significant because disorders are often unavoidable in realistic systems.

\subsection{Derivation}

We establish the proposal Eq.~(\ref{eq:dos}) in a few steps. We begin with Szegő's limit theorem for the determinant of a large Toeplitz matrix. Before doing so, it is appropriate here to introduce the terminology of Toeplitz matrices \cite{bottcher1999introduction,bottcher2005spectral}. A matrix $A$ determined by $A_{ij}=a_{j-i}$ is called a Toeplitz matrix; i.e., the matrix element $A_{ij}$ depends on the difference $j-i$ only. It is associated with a symbol, which is a complex-valued function $\sigma(e^{i\theta}) = \sum_n a_n e^{in\theta}$, $\theta\in[0,2\pi]$. A Toeplitz matrix is often expressed in terms of its symbol as $A=\mathcal{T}[\sigma]$. By definition, the elements of a Toeplitz matrix are the Fourier components of the symbol:
\begin{equation}
\mathcal{T}[\sigma]_{j_1j_2} =\int_0^{2\pi}\frac{d\theta}{2\pi}\,\sigma(e^{i\theta})e^{i(j_1-j_2)\theta}  .
\end{equation}

The language of Toeplitz matrices is very useful in addressing tight-binding Hamiltonians. For example, a single-band real-space OBC Hamiltonian $H$ in 1D is a Toeplitz matrix, and the Bloch Hamiltonian is its symbol; conversely, we say $H$ is generated by the Bloch Hamiltonian. In addition, one can generalize the series $\left\{ a_n \right\}$ from numbers to square matrices. Such a matrix $A$ is called a block Toeplitz matrix, which corresponds to a multiband Hamiltonian. One can also generalize the indices $i,j,\dots$ from integers to integer tuples with several components; for example, in 2D we take $\bm{i}=(i_x,i_y)$. The corresponding matrix $A$ is called a multilevel Toeplitz matrix, which can be viewed as the real-space Hamiltonian of a higher-dimensional lattice model. For simplicity, we call them all Toeplitz matrices hereafter.

Szegő's limit theorem was originally established for 1D Hermitian Toeplitz matrices \cite{Szego1915}, but thereafter generalized by Widom \textit{et al.} to multiband \cite{WIDOM1974284,WIDOM19761} and higher-dimensional \cite{Widom1980182,Doktorskii1984} models. To state the theorem, let us consider a subspace $\Omega$ of the $d$-dimensional Euclidean space. Let
\begin{align}
\sigma(e^{i\bm{\theta}})=\sum_{\bm{n}}t_{\bm{n}}e^{i\bm{n}\cdot\bm{\theta}} = \sum_{n_1,\dots,n_d}t_{n_1\dots n_d}e^{i(n_1\theta_1+\dots +n_d\theta_d)} \nn
\end{align}
be a symbol, which generates the Toeplitz matrix $\mathcal{T}[\sigma]$ in $\Omega$. For example, if we take $\Omega$ to be the $d$-dimensional sphere with diameter $L$ (radius $R=L/2$), then each block (or element, in the single-band cases) of $\mathcal{T}[\sigma]$ is $\mathcal{T}[\sigma]_{\bm{i}\bm{j}} =t_{\bm{j}-\bm{i}}$, with $\bm{i}=(i_1,\dots,i_d)$ and $\bm{j}=(j_1,\dots,j_d)$ satisfying $(i_1)^2+\dots+(i_d)^2\leq R^2$ and $(j_1)^2+\dots+(j_d)^2\leq R^2$. Szegő's limit theorem reveals the asymptotic behavior of the Toeplitz determinant in the $L\to\infty$ limit:
\begin{align}    \label{eq:szego}
    \log\det\mathcal{T}[\sigma]
    = N \int_{T^d} \left( \dfrac{d\theta}{2\pi} \right)^d \log\det\sigma\left( e^{i\bm{\theta}} \right)
     + O\left( L^{d-1} \right) ,
\end{align} 
where $N$ is the number of lattice points in $\Omega$, which is proportional to $L^d$ ($L$ stands for the linear size), and $T^d=[0,2\pi]^d$ is the $d$-dimensional torus. There are two conditions for Eq.~(\ref{eq:szego}) to hold: (i) $\sigma(e^{i\bm{\theta}})$ must be invertible for any $\bm{\theta}$ on the torus $T^d$, i.e., $\det\sigma(e^{i\bm{\theta}})\neq 0$. (ii) The winding number of the phase of $\det\sigma(e^{i\bm{\theta}})$ along any circle in $T^d$ must be zero, so that a matrix logarithm for $\sigma(e^{i\bm{\theta}})$ is well defined. A heuristic proof of Szegő's limit theorem is available in Appendix \ref{apd:szego}. Notably, when $\sigma(e^{i\bm{\theta}})$ and $\mathcal{T}[\sigma]$ are Hermitian, this theorem is consistent with the fact that the OBC bulk spectrum is asymptotically the same as the PBC spectrum. In fact, the left-hand side of Eq. (\ref{eq:szego}) yields the logarithm of the product of all OBC eigenvalues of $\mathcal{T}[\sigma]$, while the right-hand side yields the PBC counterpart, and these two quantities should be almost equal.  Szegő's limit theorem tells us that a similar relation remains valid under the aforementioned conditions even though $\mathcal{T}[\sigma]$ is non-Hermitian.

Now we apply this theorem to our non-Hermitian problem. Specifically, we consider $\sigma(e^{i\bm{k}})=E-h(e^{i\bm{k}})$,  then the generated Toeplitz matrix is $\mathcal{T}[E-h(e^{i\bm{k}})] =E-H$, in which $H=\mathcal{T}[h(e^{i\bm{k}})]$ is the real-space Hamiltonian corresponding to the Bloch Hamiltonian $h(e^{i\bm{k}})$. The motivation to consider $E-H$ is the simple identity
\begin{equation}
    \Phi(E)= \frac{1}{N}\log|\det(E-H)|,
\end{equation}
in which $\Phi(E)$ is the Coulomb potential defined in Eq. (\ref{coulomb}). We observe that this expression bears resemblance to the left-hand side of  Eq.~(\ref{eq:szego}), which hints useful formulas for the Coulomb potential. Before exploiting Eq.~(\ref{eq:szego}), however, we notice that its application relies on the aforementioned two conditions. For example, it would be troublesome if $\det[E-h(e^{i\bm{k}})]=0$ and therefore the symbol is not invertible at certain $\bm{k}$ points. To use Eq.~(\ref{eq:szego}), we perform a similarity transformation $D^{-1}HD$ to the Hamiltonian, with $D_{\bm{x},\bm{y}} = \delta_{\bm{x},\bm{y}}e^{\bm{\mu}\cdot\bm{x}}$. It adds an $e^{\bm{\mu}\cdot(\bm{y}-\bm{x})}$ factor to $H_{\bm{x},\bm{y}}$,  the hopping from $\bm{y}$ to $\bm{x}$, i.e., $(D^{-1}HD)_{\bm{x},\bm{y}}= H_{\bm{x},\bm{y}}e^{\bm{\mu}\cdot(\bm{y}-\bm{x})}$. As such, the corresponding Bloch Hamiltonian $h ( e^{i\bm{k}} )$ transforms into $h( e^{\bm{\mu}+i\bm{k}} )$. In the language of Toeplitz matrices, we have $D^{-1}HD=\mathcal{T}[h(e^{\bm{\mu}+i\bm{k}})]$. In fact, given $h(e^{i\bm{k}})= \sum_{\bm{n}}  t_{\bm{n}}  e^{i\bm{k}\cdot\bm{n}}$, we have $h(e^{\bm{\mu}+i\bm{k}})  = \sum_{\bm{n}}  t_{\bm{n}} e^{\bm{\mu}\cdot\bm{n}} e^{ i\bm{k} \cdot\bm{n}}$, from which we can read that $\mathcal{T}[h(e^{\bm{\mu}+i\bm{k}})]_{\bm{x},\bm{y}}  =   t_{\bm{y}-\bm{x}} e^{\bm{\mu}\cdot(\bm{y}-\bm{x})} =H_{\bm{x},\bm{y}} e^{\bm{\mu}\cdot(\bm{y}-\bm{x})}$. The transformation of $E-H$ can be written as
\begin{equation}    \label{eq:simil}
    D^{-1} \mathcal{T} \left[ E - h\left( e^{i\bm{k}} \right) \right] D
    = \mathcal{T} \left[ E - h\left( e^{\bm{\mu}+i\bm{k}} \right) \right],
\end{equation}
or $D^{-1} \mathcal{T} [ E - h ( e^{i\bm{k}}  ) ]D=\mathcal{T}  [ E - h(\bm{\beta})]$ with $\bm{\beta}= e^{\bm{\mu}+i\bm{k}}$.
It follows from Eq.~\eqref{eq:simil} that the matrix $\mathcal{T}[E-h(\bm{\beta})]$ has the same spectrum as $\mathcal{T} [ E - h(e^{i\bm{k}})]$ for an arbitrary value of $\bm{\mu}$. Therefore, we have $\det(E-H)= \det \mathcal{T}[E-h(\bm{\beta})]$ regardless of the value of $\bm{\mu}$. We can freely choose $\bm{\mu}$ such that $\sigma(\bm{\beta})=E-h(\bm{\beta})$ satisfies the conditions required in applying Eq.~\eqref{eq:szego}, even if the original symbol $E-h(e^{i\bm{k}})$ (with $\bm{\mu}=\bm{0}$) does not. In fact, when $\bm{\mu}$ locates in the central hole of the amoeba of $\det(E-h)$, we have $\det [E-h(\bm{\beta})]\neq 0$, and the winding number of $\det [E-h(\bm{\beta})]$ along any $k_j$ circle is $0$ (recall the mathematical preparation in Sec.~\ref{sec:math}). Thus, we can apply Eq.~\eqref{eq:szego} to $\sigma (\bm{\beta})=E-h(\bm{\beta})$ and take the real part, resulting in
\begin{align}    \label{sigmabeta}
    &\phantom{=} \frac{1}{N}\log|\det\mathcal{T}[E-h(\bm{\beta})]|  \notag\\
    &= \int_{T^d} \left( \dfrac{dk}{2\pi} \right)^d \log|\det[E-h\left( \bm{\beta} \right)]|  + O\left( L^{-1} \right),
\end{align}
where $\bm{\beta}= e^{\bm{\mu}+i\bm{k}}$ with $\bm{\mu}$ fixed in the central hole of the amoeba of $E-h(\bm{\beta})$. As has been explained, the left-hand side of Eq.~(\ref{sigmabeta}) is $\frac{1}{N}\log\left| \det \left( E-H \right) \right|$. Notably, the integral on the right-hand side is exactly the Ronkin function of $\det(E-h)$. Thus,  Eq. (\ref{sigmabeta}) can be written as
\begin{equation}    \label{PhiR}
     \dfrac{1}{N}\log\left| \det \left( E-H \right) \right| = R_{\det(E-h)} (\bm{\mu}) +  O\left( L^{-1} \right),
\end{equation}
in which $\bm{\mu}$ locates in the central hole of the amoeba. Since the Ronkin function takes its minimum in the central hole, Eq. (\ref{PhiR}) reduces to Eq. (\ref{Phiphi}) in the $N\to\infty$ limit.

So far, this proof of Eq. (\ref{Phiphi}) is incomplete because the existence of the central hole of the amoeba is assumed. When the central hole does not exist, Szegő's limit theorem Eq.~\eqref{eq:szego} cannot be applied in its original form. Here, we propose a generalization of Eq.~\eqref{eq:szego} to remove this limitation. The generalization makes essential use of the Ronkin function. The proposed generalization is
\begin{equation}    \label{szego-generalized}
    \log\det\mathcal{T}[\sigma]
    = N \int_{T^d} \left( \dfrac{dk}{2\pi} \right)^d \log\det\sigma\left( \bm{\beta} \right)
     + O\left( L^{d-1} \right),
\end{equation}
in which $\bm{\beta}= e^{\bm{\mu}_\text{min}+i\bm{k}}$ with $\bm{\mu}_\text{min}$ being the minimum location of the Ronkin function; i.e., $R_{\det\sigma} (\bm{\mu}) $ takes its minimum at $\bm{\mu}_\text{min}$. Note that the left-hand side of Eq.~(\ref{szego-generalized}) can be taken as  $\log\det\mathcal{T}[\sigma]=\log\det\mathcal{T}[\sigma(e^{\bm{\mu}+i\bm{k}})]$ with an arbitrary $\bm{\mu}$. This is because a similarity transformation of the real-space Hamiltonian does not change its determinant [see the discussion below Eq.~\eqref{eq:simil}]. The location $\bm{\mu}_\text{min}$ can be determined by the vanishing of gradient $\partial_{\mu_j}  R_{\det\sigma}  =0$ for all $j$'s (or simply written as $\partial_{\bm{\mu}}  R_{\det\sigma} = \bm{0}$).

\begin{figure}[t]
    \centering
    \includegraphics[width=\linewidth]{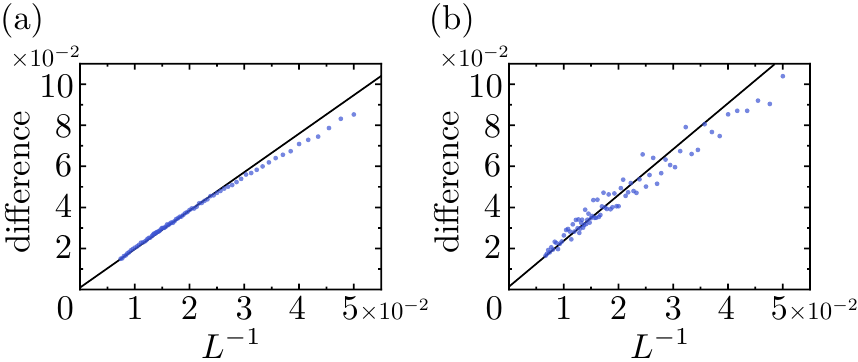}
    \caption{The maximal difference $\max_E|\phi(E)-\Phi(E)|$ between the OBC-spectrum-based and Ronkin-function-based Coulomb potentials as a function of the linear size $L$. The adopted Hamiltonian is Eq. \eqref{eq:dgmodel} [Fig.~\ref{fig:hopping}(a)]. The straight line is a linear fitting from data points with $L\geq 55$. (a) The real space is a square of side length $L$, with disorders added on the boundary [the same as in Fig.~\ref{fig:dos}(b)]. Each data point comes from averaging over six disorder configurations. (b) The real space is a disk of diameter $L$, without disorder [the same as in Fig.~\ref{fig:dos}(c)].}
    \label{fig:scaling}
\end{figure}

When the amoeba of $\det\sigma(\bm{\beta})$ contains a central hole, Eq. (\ref{szego-generalized}) can be established by the same approach that leads to Eq. (\ref{PhiR}). When there is no central hole, we may intuitively view the minimum location $\bm{\mu}_\text{min}$ as an ``infinitesimal central hole,'' which is consistent with the fact that the Ronkin function takes the minimum in the central hole. Although we do not find a rigorous proof of Eq. (\ref{szego-generalized}) in these general cases, our numerical results support its validity (see below). In fact, taking $\sigma(\bm{\beta})=E-h(\bm{\beta})$ in Eq. (\ref{szego-generalized}) and extracting the real part lead to \begin{align}
     \dfrac{1}{N}\log\left| \det \left( E-H \right) \right| = R_{\det(E-h)} (\bm{\mu}_\text{min})+  O\left( L^{-1} \right),
\end{align} or \begin{align}    \label{PhiphiL}
     \Phi(E) = \phi(E) +  O\left( L^{-1} \right),
\end{align} which becomes Eq. (\ref{Phiphi}) in the large-size limit $L\rightarrow\infty$. To confirm this, we numerically calculate the Coulomb potential $\Phi(E)$ for different system sizes, and compare it with the Ronkin minimum $\phi(E)$.  In Fig.~\ref{fig:scaling}, we plot the maximal difference $\max_E|\phi(E)-\Phi(E)|$ as a function of the system size $L$. Note that the size dependence comes solely from that of $\Phi(E)$, while $\phi(E)$ is independent of the size. Regardless of the shape of the OBC system, the maximal difference is in good agreement with the $L^{-1}$ behavior, and converges to $0$ when extrapolated to the $L\to\infty$ limit. These behaviors are exactly what Eq.~(\ref{PhiphiL}) tells.

For Hermitian bands, the DOS of OBC systems determined by our formulation using the Ronkin function is consistent with the familiar fact that the OBC bulk spectrum and the PBC spectrum are asymptotically identical. This can be proved as follows. Since the eigenenergies belong to the real axis $\mathbb{R}$, one can determine the DOS by Eq. (\ref{1D}), for which knowing $\phi(E)$ for $E\notin\mathbb{R}$ is sufficient. For an arbitrary $E\notin\mathbb{R}$, $\det[E-h(e^{i\bm{k}})]\neq 0$ always holds for all $\bm{k}\in[0,2\pi]^d$ because of the Hermiticity of $h(e^{i\bm{k}})$. Thus, $\bm{\mu}=\bm{0}$ does not belong to the amoeba of $\det[E-h(\bm{\beta})]$. In other words, it belongs to one of the amoeba holes. Furthermore, the phase winding number of $\det[E-h(e^{i\bm{k}})]$ along every $k_j$ circle is zero. Therefore, the order $\nu_j=0$ ($j=1,\dots , d$) at $\bm{\mu}=\bm{0}$, meaning that $\bm{\mu}=\bm{0}$ locates in the central hole.  It follows that the Ronkin minimum is reached at $\bm{\mu}=\bm{0}$, i.e., $\phi(E)=R_{\det(E-h)}(\bm{0})$. On the other hand, one can readily see that $R_{\det(E-h)}(\bm{0})$ by definition is the Coulomb potential of the PBC energy spectrum. Therefore, the OBC DOS obtained from Eq. (\ref{1D}) is the same as the PBC DOS in Hermitian cases.

\section{\label{sec:boundary}Amoeba hole closing and spectral boundary}

We are now able to prove a powerful theorem about the range of the OBC spectrum in the complex energy plane. It uses only the topology of the amoeba, without having to evaluate the Ronkin function. Hence, it is more efficient when one only wants to know the range of the spectrum. 

We denote by $\Lambda$ the set of $E$ where the amoeba of $\det[E-h(\bm{\beta})]$ does not possess a central hole. We prove the following theorem:
\begin{equation} \label{rhozero}
    \rho(E) = 0, \quad
    E\notin\Lambda.
\end{equation}
Thus, at a certain $E$, if the amoeba has a central hole, the DOS is zero at this $E$. In other words, the bulk spectrum is restricted inside $\Lambda$.

\begin{figure}[t]
    \centering
  \includegraphics[width=\linewidth]{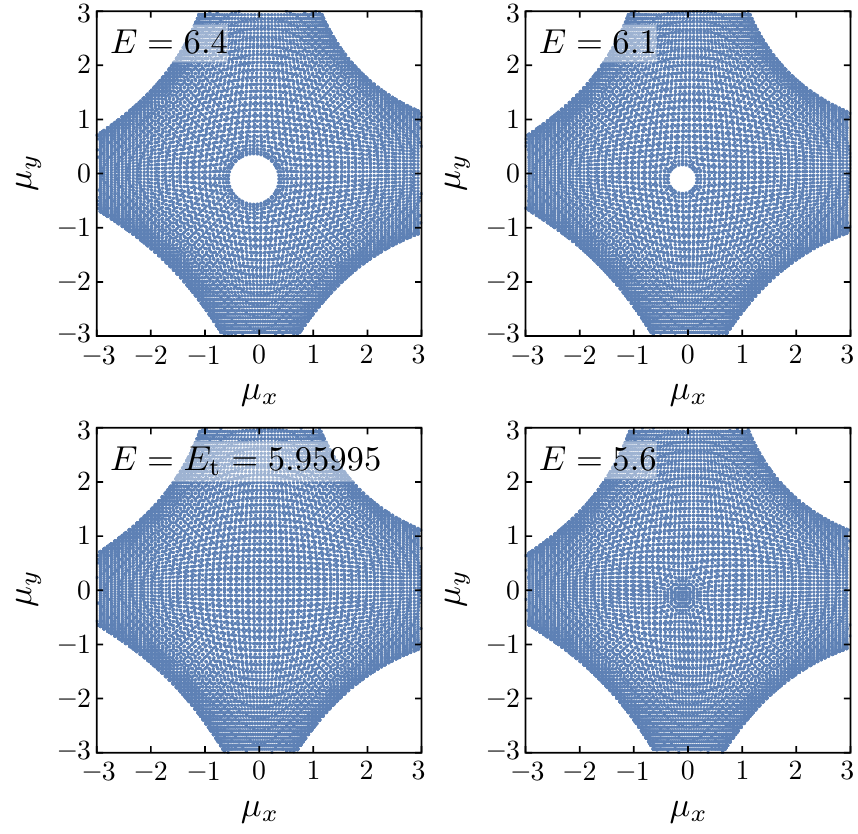}
    \caption{Amoebae for several energies near the band top of the model Eq.~\eqref{eq:dgmodel}. As we decrease the energy along the real axis, the central hole of amoeba closes at $E=E_{\mathrm{t}}\approx \num{5.95995}$. Parameter values are $t=1$, $t'=0.5$, and $\gamma=0.2$, for which the energy spectrum on a disk is shown in Fig. \ref{fig:amoeba}(d). }
    \label{fig:vacuole}
\end{figure}

With the results of the DOS in the previous section, the proof of this theorem is now simple. When an energy $E_0$ is outside $\Lambda$, one must choose $\bm{\mu}$ in the central hole so that Eq.~(\ref{PhiR}) holds. Because the shape of the central hole varies continuously as $E$ varies, the central hole should still contain this $\bm{\mu}$ for $E$ sufficiently close to $E_0$.   Therefore, there exists a neighborhood of $E_0$ denoted by $V$, such that for any $E\in V$, the same $\bm{\mu}$ is in the central hole of the amoeba of $\det(E-h)$. Thus, for any $E\in V$, the Ronkin minimum is $\phi(E)=R_{\det(E-h)} (\bm{\mu})$. Recalling the definition of amoeba, we see that $\det[E-h(e^{\bm{\mu}+i\bm{k}})]$ is nowhere zero in $V$ for any $\bm{k}\in T^d=[0,2\pi]^d$. Furthermore, the Ronkin function can be expressed as
\begin{equation}
    R_{\det(E-h)} (\bm{\mu})=\int_{T^d} \left( \frac{dk}{2\pi} \right)^d \sum_i \log | E-E_i(\bm{\beta})|   ,
\end{equation}
in which $\{E_i(\bm{\beta})\}$ are the eigenvalues of $h(\bm{\beta})$ ($\bm{\beta}=e^{\bm{\mu}+i\bm{k}}$ as usual). Because $\det(E-h)\neq 0$ implies $|E-E_i(\bm{\beta})|\neq 0$, we have $\Delta[\sum_i\ln|E-E_i(\bm{\beta})|]=0$ and therefore its integration over $T^d$ vanishes, leading to $\rho(E)=\Delta\phi(E)/2\pi = 0$. This ends our proof of the theorem Eq. (\ref{rhozero}). Note that this proof makes use only of the original version of Szegő's theorem Eq.~\eqref{eq:szego}, without invoking the generalized version Eq.~\eqref{szego-generalized}.

When $E\in\Lambda$, the DOS is generally nonzero since nothing forces it to vanish. Thus, the boundary of $\Lambda$, on which the central hole closes, coincides with the boundary of the energy spectrum. In fact, if we assume the validity of the conjecture Eq.~(\ref{szego-generalized}) and therefore Eq.~\eqref{Phiphi}, we have $\rho(E)=\Delta\phi(E)/2\pi$, which is generally nonzero in $\Lambda$. In other words, the support of the DOS is exactly $\Lambda$.

In Fig.~\ref{fig:vacuole}, we illustrate the amoeba-hole closing for the model Eq.~\eqref{eq:dgmodel}. Starting from an energy well above the band top (maximum of the real part of the eigenenergies),  we decrease the energy along the real axis. For $E$ larger than a certain energy, the central hole of amoeba exists, though its size shrinks as $E$ decreases. At $E=E_{\mathrm{t}}\approx \num{5.95995}$, the central hole closes. According to our proposal, this hole-closing point is identified as the band top. Similarly, the entire spectral boundary can be delineated by the hole closing.

\begin{figure}[t]
    \centering
    \includegraphics[width=\linewidth]{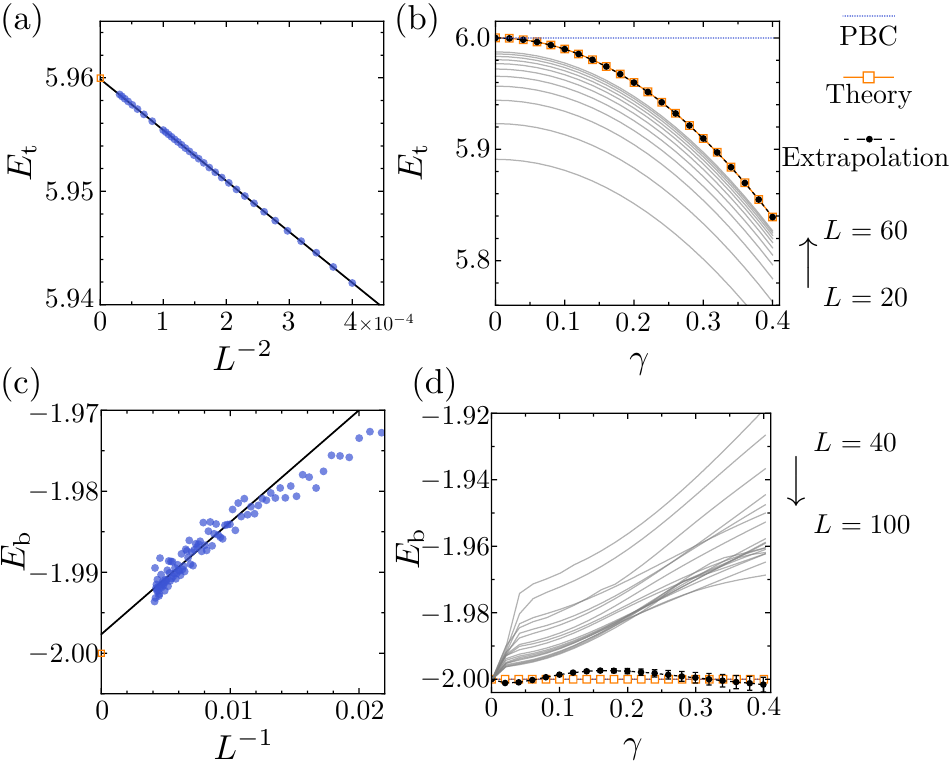}
    \caption{Band top $E_{\mathrm{t}}$ and bottom $E_{\mathrm{b}}$ obtained from amoeba formulation and numerical calculations. The model is Eq.~\eqref{eq:dgmodel}, with $t=1$, $t'=0.5$ fixed. In (a)(c), $\gamma$ is fixed to $0.2$. (a) Band top obtained from diagonalizing the real-space Hamiltonian on the disk with diameter $L$. The extrapolation to $L\rightarrow\infty$ agrees well with the amoeba-hole-closing point marked as the orange square. (b) Band top as a function of $\gamma$, for increasing $L$ values. The extrapolation to $L\rightarrow\infty$ is shown as black dots, which are in excellent agreement with the amoeba-hole-closing points marked as orange squares. The PBC result is shown as the dotted line. (c)(d) The counterparts of (a)(b) for the band bottom. The linear fitting in (c)(d) is based on data from $L\in[80,240]$. Error bars indicate $95\%$ confidence intervals. }
    \label{fig:topbottom}
\end{figure}

In Fig.~\ref{fig:topbottom}, we compare the results obtained from the amoeba formulation and numerical calculations. We numerically diagonalize the real-space Hamiltonian with increasing sizes and obtain the values of the band top, which are then extrapolated to infinite size [Fig.~\ref{fig:topbottom} (a)]. The results are strikingly close to the predictions of the amoeba theory [Figs.~\ref{fig:topbottom} (a)(b)]. We also do a similar comparison for the band bottom (minimum of the real part of the eigenenergies) [Figs.~\ref{fig:topbottom} (c)(d)].  Note that the band top and band bottom exhibit different scaling behavior when $L\to\infty$: the finite-size correction $\Delta E_{\mathrm{t}}  \propto L^{-2}$ for the former [Fig.~\ref{fig:topbottom} (a)], and $\Delta E_{\mathrm{b}} \propto L^{-1}$ for the latter [Fig.~\ref{fig:topbottom} (c)]. It turns out that the former scaling is more accurately obeyed here. Therefore, the error of extrapolation is larger for the latter. Despite larger numerical error, the numerical results are still in good agreement with the amoeba-theoretic prediction.

\section{\label{sec:gbz}Generalized Brillouin zone}


In this section, we establish another proposal mentioned at the end of Sec.~\ref{sec:math}, namely, the location of the Ronkin minimum determines the complex momenta and hence the GBZ. According to Eq. (\ref{rhozero}), the DOS can be nonzero only when the central hole of the amoeba of $\det(E-h)$ is absent. Thanks to the convexity of the Ronkin function, the minimum location $\bm{\mu}_{\min}$ must be unique in these cases, and therefore the proposal is unambiguous.

\begin{figure}[t]
    \centering
    \includegraphics[width=\linewidth]{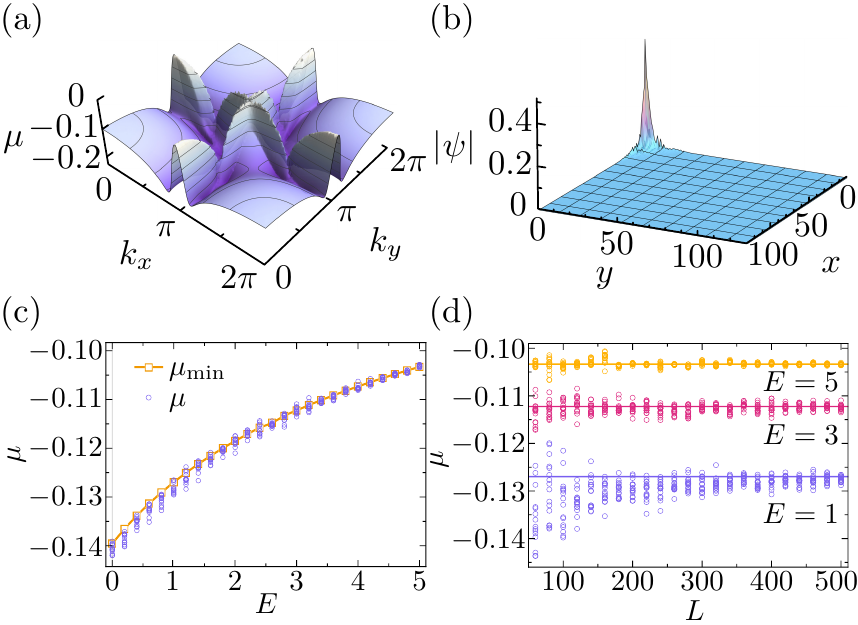}
    \caption{(a) The generalized Brillouin zone of model Eq.~(\ref{eq:dgmodel}). Because of the symmetry of interchanging $x$ and $y$ in this model, we have $(\bm{\mu}_\mathrm{min})_x = (\bm{\mu}_\mathrm{min})_y\equiv\mu$, which is plotted as a function of the real part of the wave vector $\bm{k}=(k_x,k_y)$. 
    (b) Typical profile of a bulk eigenstate that exhibits the non-Hermitian skin effect. The eigenstate is taken at $E=-0.003+0.056i$. The system is a square of length $L=130$, with certain random on-site disorders on the boundary. (c)(d) Comparison between $\mu_{\mathrm{min}}$ and the exponential decay rate of the Green's function.  The latter is defined by linearly fitting $\mu_x,\mu_y$ from $\log\bra{\bm{x}} (E-H)^{-1} \ket{\bm{0}}\sim  \mu_x x+\mu_y y$, in which $\mu_x\approx\mu_y$ for our model and therefore we present the average $\mu=(\mu_x+\mu_y)/2$. Here, $H$ is a real-space Hamiltonian defined on a disk with diameter $L$, with random on-site disorder (distributed uniformly in $[-0.5, 0.5]$) on the boundary. Each circle represents the result from a disorder configuration. In the fitting, we discard $\bm{x}$'s that are within distance $20$ from the boundary. In (c), we fix $L=400$. In (d), we vary $L$ at fixed $E=1,3,5$; the corresponding $\mu_\text{min}$'s are shown as horizontal lines. }
    \label{fig:gbz}
\end{figure}

The GBZ essentially determines the exponential behavior of the eigenstates of a non-Hermitian lattice Hamiltonian. According to our proposal, an eigenstate with eigenenergy $E$ is expressed asymptotically in the bulk as \footnote{Note that possible boundary terms are omitted, in the same spirit as in 1D.}
\begin{equation} \label{eigenstate}
    \psi_E (\bm{x}) = \sum_{\bm{k}} c_{\bm{k}} \exp \left[ \left( i\bm{k}+ \bm{\mu}_{\min} \left( E \right)  \right) \cdot \bm{x} \right],
\end{equation}
where the sum is over all $\bm{k}$ satisfying $\det[E-h( e^{\bm{\mu}_\text{min}+i\bm{k}} )]=0$, and $c_{\bm{k}}$ are certain $E$-dependent coefficients. Here, $\bm{\mu}_{\min}$ plays the role of the imaginary part of the wave vector if we define a complex-valued wave vector
\begin{equation}
    \tilde{\bm{k}} = \bm{k} - i\bm{\mu}_{\min}.
\end{equation}
Equivalently, we can use the variable $\bm{\beta}=e^{i\tilde{\bm{k}}}$ (or $\tilde{\bm{k}} = -i\log\bm{\beta}$), so that $\psi_E (\bm{x}) = \sum_{\bm{\beta}} c_{\bm{\beta}} \prod_{j=1}^d(\beta_j)^{x_j}$. In our theory, the GBZ consists of all points $\bm{\beta}$ subject to
\begin{equation}    \label{gbzeq}
    \begin{aligned}
    \det[E-h(\bm{\beta})] &= 0,  \\
     (\log|\beta_1|,\dots,\log|\beta_d|) &= \bm{\mu}_{\min}(E). \end{aligned}
\end{equation}
For a $d$-dimensional non-Hermitian system, the GBZ is a $d$-dimensional subspace of the $\bm{\beta}$ space whose real dimension is $2d$. In fact, there are $2d+2$ real unknowns in Eq. (\ref{gbzeq}), namely, the real and imaginary parts of $(\bm{\beta},E)$. Equation (\ref{gbzeq}) then imposes $d+2$ constraints, meaning that the solution space is $d$ dimensional. Equation (\ref{gbzeq}) provides a general approach to calculate the GBZ for higher-dimensional non-Hermitian systems. In practice, we often parametrize the GBZ by $\bm{k}$, treating $\bm{\mu}_{\min}$ as its function. This vectorial function $\bm{\mu}_{\min}(\bm{k})$ is a complete representation of the GBZ. As an application of our theory, the GBZ thus obtained for our model Eq.~\eqref{eq:dgmodel} is shown in Fig.~\ref{fig:gbz}(a).


Now we provide evidence for Eqs. (\ref{eigenstate}) and (\ref{gbzeq}). First, we show that an eigenstate $|\psi\rangle$ subject to $H|\psi\rangle = E|\psi\rangle$ shares the same exponential behavior with the Green's function $G(E) = (E-H)^{-1}$. To see this, we decompose the OBC Hamiltonian into its eigenstates:
\begin{equation}
    H = \sum_{n} \epsilon_n \ket{nR} \bra{nL}  ,
\end{equation}
where the sum is over all eigenstates; $\ket{nR}$ and $\bra{nL}$ are the right and left eigenstates, respectively, which satisfy
\begin{equation}
    H \ket{nR} = \epsilon_n \ket{nR}    ,   \quad
    \bra{nL} H = \epsilon_n \bra{nL}    .
\end{equation}
They are orthonormalized as $\bracket{mL}{nR} = \delta_{mn}$. Suppose that $\bm{x}$ and $\bm{0}$ are real-space locations far from the boundary. The Green's function is related to the eigenstates in the following way:
\begin{equation}    \label{eq:green}
    \begin{aligned}
        &\phantom{=} \oint_{C} \frac{dE'}{2\pi i} \bra{\bm{x}} \left( E'-H \right)^{-1} \ket{\bm{0}} \\ 
        &= \sum_{n} \bracket{\bm{x}}{nR} \bracket{nL}{\bm{0}} \oint_{C} \frac{dE'}{2\pi i} \frac{1}{E'-\epsilon_{n}}  \\
        &= \sum_{\epsilon_n = E} \bracket{nL}{\bm{0}} \bracket{\bm{x}}{nR}  ,
    \end{aligned}
\end{equation}
where $C$ is an infinitesimal counterclockwise contour centered at $E$. We emphasize that $E$ is taken in the OBC energy spectrum, otherwise the integral in Eq.~\eqref{eq:green} would vanish. The right-hand side of Eq.~\eqref{eq:green} is a linear combination of all eigenstates with energy $E$. As our objective is to obtain $\bm{\mu}$, it suffices to keep both the eigenstates and the Green's functions to the exponential level. Assuming that the Green's function does not change abruptly with respect to $E$, we conclude that these eigenstates share the same exponential behavior as the Green's function $\bra{\bm{x}} \left( E-H \right)^{-1} \ket{\bm{0}}$. We would also like to remark that, while the above argument assumes that the Hamiltonian is diagonalizable, the result is general enough to hold even in the presence of exceptional points \footnote{In the presence of EPs, we have an additional nilpotent term $N$ in the spectral decomposition: $H = \sum_n \epsilon_n \ket{nR}\bra{nL} + N$. The Green's function also acquires an additional term, roughly as $(E-H)^{-1} = \sum_n \ket{nR}\bra{nL}/(E-\epsilon_n) + \sum (\textrm{nilpotent matrix})/(E-\epsilon_n)^p$, where $p\geq 2$ \cite{Ashida2021}. The final term vanishes after the contour integral in Eq.~\eqref{eq:green}, and therefore Eq.~\eqref{eq:green} still holds. }. 

Suppose that the asymptotic behavior is $\bra{\bm{x}} \left( E-H \right)^{-1} \ket{\bm{0}}\sim \exp(\sum_{j=1}^d \mu_j x_j)$. To validate Eqs. (\ref{eigenstate}) and (\ref{gbzeq}), we need to show that $\mu_j =(\mu_\text{min})_j$. We numerically confirm this relation in Figs. \ref{fig:gbz}(c) and \ref{fig:gbz}(d). Note that in principle one may also validate Eqs. (\ref{eigenstate}) and (\ref{gbzeq}) by the exponential behavior of eigenstates themselves. However, our detour of calculating the Green's function is computationally more accurate and less expensive, enabling calculation for larger $L$ within reasonable time.

\section{\label{sec:topo}Non-Bloch band topology}

It has been found recently that the topological numbers defined on the conventional Brillouin zone fail to characterize the topological edge states and bulk-boundary correspondence in non-Hermitian systems. To correctly account for the topological edge modes, the non-Bloch topological invariants defined on the GBZ have been proposed. In practice, most of their applications are restricted to 1D, because a general calculable formulation of GBZ in higher dimension has been lacking. Based on the amoeba formulation, we are now able to address the non-Bloch band topology in higher dimensions. Specifically, the non-Bloch Chern number, which was previously calculated only by continuum approximation, can now be calculated in the entire GBZ as is.


\begin{figure}[t]
    \centering
    \includegraphics[width=\linewidth]{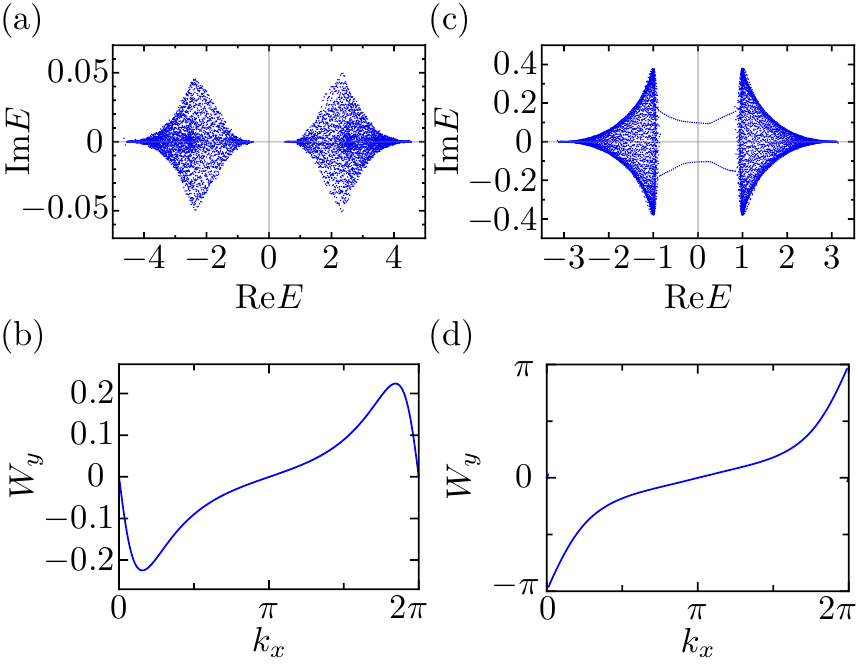}
    \caption{Energy spectrum and band topology of the non-Hermitian Chern-band model Eq.~\eqref{eq:qwz}. (a) Energy spectrum by diagonalizing the real-space Hamiltonian on a square of side length $L=60$, with random on-site disorders on boundary. Parameters are taken in the topologically trivial regime: $m=2.6$, $\gamma=0.4$. (b) The Berry phase $W_y(k_x)$ along a circle in the $k_y$ direction ($k_x=\mathrm{const}$). The non-Bloch Chern number $C=[W_y(2\pi)-W_y(0)]/2\pi=0$. Parameter values are the same as in (a). (c) Energy spectrum by diagonalization, calculated under the same setting as in (a). Parameters are taken in the topologically nontrivial regime: $m=1.2$, $\gamma=0.4$. Topological edge states are observed. (d) The Berry phase as a function of $k_x$, with $W_y(2\pi)-W_y(0)=2\pi$ and therefore $C=1$.  Parameter values are the same as in (c). }
    \label{fig:chern}
\end{figure}

To be concrete, we consider the following non-Hermitian Chern-band model \cite{yao2018chern}
\begin{align} \label{eq:qwz}
    h \left( e^{i\bm{k}} \right) &= \left( v \sin k_x + i\gamma \right) \sigma_x
    + \left( v \sin k_y + i\gamma \right) \sigma_y    \notag\\
    &\phantom{=} + \left( m - t \cos k_x - t \cos k_y \right) \sigma_z  .
\end{align}
The real-space Hamiltonian reads
\begin{align}
    H &= \sum_{\bm{x}}\sum_{j=x,y} \ket{\bm{x}} \left( -\dfrac{i}{2}v\sigma_j - \dfrac{1}{2}t\sigma_z \right) \bra{\bm{x}+\bm{e}_j} + \mathrm{H.c.}
    \notag\\
    &\phantom{=} + \sum_{\bm{x}} \ket{\bm{x}} \left( m\sigma_z + i\gamma\sigma_x + i\gamma\sigma_y \right) \bra{\bm{x}}   ,
\end{align}
where $\bm{x}$ are 2D integer coordinates, and $\bm{e}_j$ is the unit vector in the $j$th direction [see Fig. \ref{fig:hopping}(b)]. We set $t=v=1$ for simplicity.



This model is known to have a Chern insulator phase as well as a trivial insulator phase in the Hermitian limit ($\gamma=0$) \cite{qi2005}. It has Chern number $C=1$ when $0<m<2$, $C=-1$ when $-2<m<0$, and $C=0$ otherwise. We focus on the phase boundary between $C=0$ and $C=1$ . When we turn on the non-Hermitian term ($\gamma \neq 0$), the phase boundary extends into a curve in the $m$-$\gamma$ plane, which should be predicted by the non-Bloch band theory. In particular, the number of chiral edge modes in each phase is given by the non-Bloch Chern number evaluated on the GBZ, which is a two-dimensional subspace of the four-dimensional $\bbeta$ space. Alternatively, we may take $\tilde{\bm{k}}=\bk-i\bm{\mu}=\log\bbeta$ as the coordinate system, in which the GBZ is two dimensional because $\bm{\mu}$ is treated as a function of $\bk$, so that the GBZ can be parametrized by $\bk$. Thus, the non-Bloch Chern number can be viewed as the conventional Chern number of $h(e^{i\tilde{\bm{k}}})$, with $\tilde{\bm{k}}\in\text{GBZ}$. For a band labeled by $\alpha$, the non-Bloch Chern number reads
\begin{equation}
    C = \dfrac{1}{2\pi i} \int_{T^2} dk_x dk_y \, \epsilon^{ij} \partial_i \bra{ \tilde{\bm{k}},\alpha, L } \partial_j \ket{ \tilde{\bm{k}},\alpha, R } ,
\end{equation}
where $\epsilon^{xy}=-\epsilon^{yx}=1$, $\partial_j$ refers to $\frac{\partial}{\partial k_j}$, and $\langle \tilde{\bm{k}},\alpha,L |$ and $| \tilde{\bm{k}},\alpha,R \rangle$ are the left and right eigenstates of the $\alpha$ band, respectively. We shall focus on the ``valence band'' ($\Re E < 0$), labeled as $\alpha=-$.

More intuitively, the non-Bloch Chern number can be expressed as the change of Berry phase along circular sections of the GBZ in the $y$ direction, i.e., $C=\frac{1}{2\pi}\int_0^{2\pi} \frac{dW_y(k_x)}{dk_x} dk_x=[W_y(2\pi)-W_y(0)]/2\pi$, in which the Berry phase
\begin{align}
    W_y \left( k_x \right) &= -i\int_0^{2\pi} \bra{\tilde{\bm{k}},-,L} \partial_{k_y} \ket{\tilde{\bm{k}},-,R} d k_y \notag\\
    &= -\int_0^{2\pi} \dfrac{h_x\partial_{k_y}h_y-h_y\partial_{k_y}h_x}{2 \bar{h} \left( \bar{h} + h_z \right)} d k_y   ,
\end{align}
where we have write the Bloch Hamiltonian as $h = h_x\sigma_x + h_y\sigma_y + h_z\sigma_z$, and $\bar{h} = \sqrt{h_x^2+h_y^2+h_z^2}$. Here, $h_{x,y,z}$ can be extracted from Eq.~\eqref{eq:qwz} by the substitution $\bk\rightarrow\tilde{\bk}$: $h_x= v \sin \tilde{k}_x + i\gamma$, $h_y = v \sin \tilde{k}_y + i\gamma$, and $h_z= m - t \cos \tilde{k}_x - t \cos \tilde{k}_y$.

In Fig.~\ref{fig:chern}, we show the energy spectrum and the non-Bloch Chern number for the Chern-band model. As an example, we fix $\gamma=0.4$, and take $m=2.6$ and $m=1.2$. From Figs.~\ref{fig:chern}(b) and \ref{fig:chern}(d), we read for each case that $C=0$ and $C=1$, respectively. In Fig.~\ref{fig:chern}(c), topological edge states are indeed seen, being consistent with the nonzero non-Bloch topological invariant $C=1$.

\begin{table}[b]
    \caption{\label{tab:1} Topological phase boundary $m_c$ as a function of $\gamma$. The decay factor $\mu$ (equal in both $x$ and $y$ directions) at $E=0$ and ``$m_c$ amoeba'' is determined using the amoeba method, with numerical error $<$ $1\times 10^{-6}$. ``$m_c$ numerical'' is adopted from Ref. \cite{yao2018chern}, which is obtained by numerical diagonalization of the real-space Hamiltonian with numerical error $<$ $3\times 10^{-4}$.   }
    \begin{ruledtabular}
        \begin{tabular}{cccc}
            $\gamma$ & $\mu$ & $m_c$ amoeba & $m_c$ numerical  \\
            \hline
            $0.00$ & $0.000000$ & $2.000000$ & $2.0000$   \\
            $0.05$ & $0.049979$ & $2.002498$ & $2.0025$    \\
            $0.10$ & $0.099834$ & $2.009975$ & $2.0100$    \\
            $0.15$ & $0.149443$ & $2.022375$ & $2.0225$    \\
            $0.20$ & $0.198690$ & $2.039608$ & $2.0400$    \\
            $0.25$ & $0.247466$ & $2.061553$ & $2.0625$    \\
            $0.30$ & $0.295673$ & $2.088061$ & $2.0885$    \\
            $0.35$ & $0.343222$ & $2.118962$ & $2.1200$    \\
            $0.40$ & $0.390035$ & $2.154066$ & $2.1540$    \\
            $0.45$ & $0.436050$ & $2.193171$ & $2.1940$    \\
            $0.50$ & $0.481212$ & $2.236068$ & $2.2360$
        \end{tabular}
    \end{ruledtabular}
\end{table}

\begin{figure}[t]
    \centering
    \includegraphics[width=\linewidth]{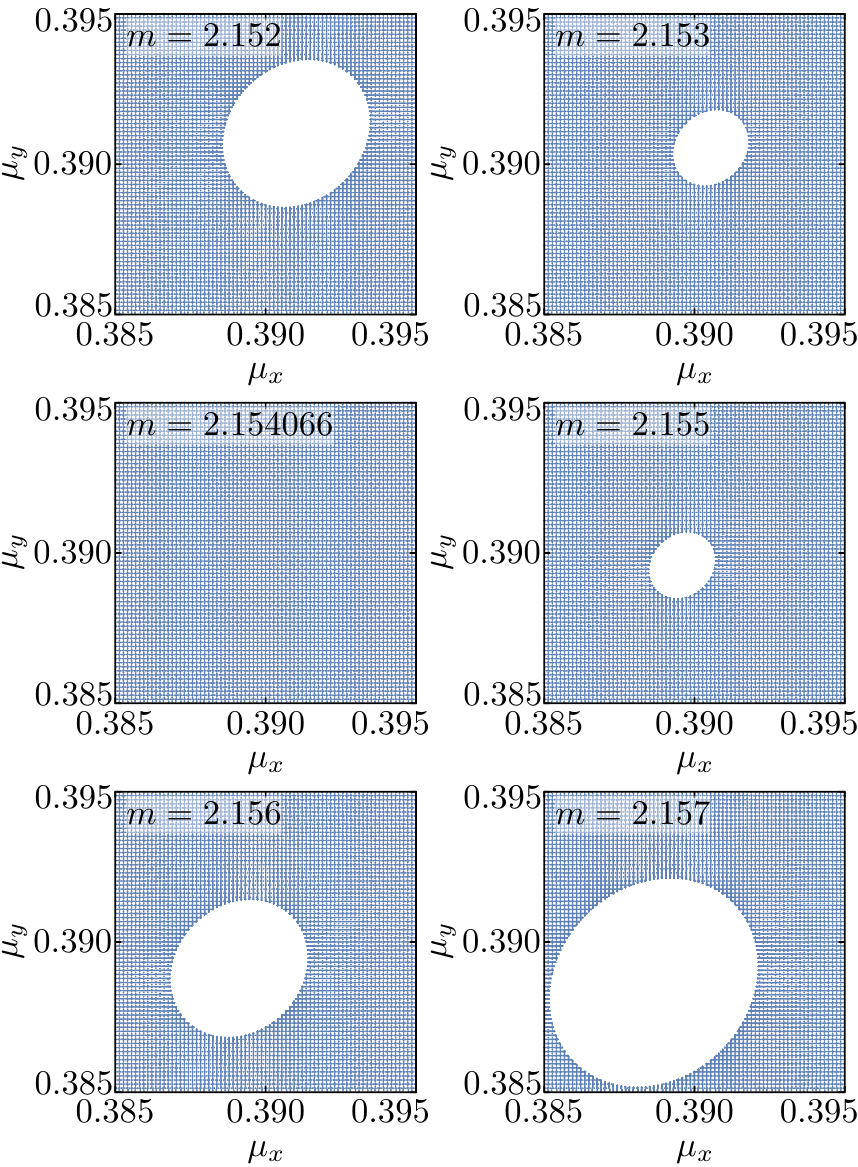}
    \caption{The amoeba-based procedure of locating the phase boundary in Table~\ref{tab:1}. The figures show the amoebae, zooming in on the central hole, with $\gamma = 0.4$ fixed and $m$ varied. The central hole closes at exactly one point $m=m_c\approx \num{2.154066}$, which is where the topological phase transition occurs. }
    \label{fig:chernamoeba}
\end{figure}


For the present model, the precise phase boundary can even be analytically determined by the amoeba formulation. In the  $m$-$\gamma$ plane, the phase boundary between the $C=0$ and $C=1$ phases is a curve, which can be viewed as a function $m_c(\gamma)$; i.e., the phase transition point is $m_c(\gamma)$ for a fixed $\gamma$.  On the phase boundary, the energy gap about $E=0$ closes, and $E=0$ belongs to the energy spectrum. Thus, we can find $m_c$ by inspecting when the central hole of the amoeba at $E=0$ closes. The exact formula for $m_c$ is found to be
\begin{equation} \label{eq:mc}
    m_c = 2 \sqrt{1+\gamma^2}   .
\end{equation}
A perturbative formula for $m_c$ was obtained in Ref. \cite{yao2018chern}, and recently a result to the sixth order of $\gamma$ was reported in Ref. \cite{takane2021bulk}.  Our derivation of Eq.~\eqref{eq:mc} is as follows. We focus on the special solution of the characteristic equation that satisfies $\beta_x = \beta_y \equiv \beta$, as the symmetry of the model allows. The characteristic equation $\det[E-h(\beta_x, \beta_y)] = 0 $ becomes
\begin{equation}
    \left( m-\beta-\beta^{-1} \right)^2 - 2 \left( \gamma - \dfrac{\beta-\beta^{-1}}{2} \right)^2 = 0   .
\end{equation}
In a neighborhood of the phase transition, we numerically notice that the four roots of the above quartic equation are all real, and that the second-largest and third-largest (the middle two) roots, under the logarithm, are on the boundary of the central hole of the amoeba. Hence requiring the central hole to vanish is equivalent to requiring the middle two roots to be equal. Solving this requirement leads to the final result Eq.~\eqref{eq:mc}.


Equation \eqref{eq:mc} is verified up to high precision by numerically locating where the central hole closes, as listed in Table \ref{tab:1}. From the amoeba-based approach, Fig.~\ref{fig:chernamoeba} illustrates locating the critical value $m_c$ for the case $\gamma=0.4$.  In practice, it is convenient to search for $m_c(\gamma)$ by iteration:
\begin{enumerate}
    \item Plot the amoebae at a consecutive sequence of $m$'s.
    \item Find the interval with the smallest central-hole area (closest to hole closing).
    \item Terminate if the desired accuracy is reached.
    \item Otherwise, on this new interval, assign a new sequence of $m$'s and go back to step 1.
\end{enumerate}
On the other hand, $m_c$ can also be determined by numerically diagonalizing the real-space Hamiltonian and finding where the band gap closes \cite{yao2018chern}. We compare the amoeba-based and numerical-diagonalization results, which agree well with each other.


In the amoeba formulation, an exact expression for the decay factor $\mu_x=\mu_y\equiv\mu$ at the band-closing point $E=0$ can also be obtained:
\begin{equation}
    \mu=\log(\sqrt{\gamma^2+1}+\gamma).
\end{equation}
This result is very close to the $\mu$ values listed in Table~\ref{tab:1}, which are numerically obtained by locating the amoeba-hole-closing points. We remark that a first-order approximation $\mu = \gamma + o(\gamma)$ was reported in Ref.~\cite{yao2018chern}.

\section{\label{sec:ineq}Spectral inequalities}

As one of the applications of the amoeba formulation, we prove a general inequality about the spectral radius. For a matrix or operator $H$, the spectral radius $\rho(H)$ is defined to be the largest absolute value of the eigenvalues. Our spectral inequality states that for a non-Hermitian lattice Hamiltonian with translational symmetry in the bulk, the OBC spectral radius is less than or equal to its PBC counterpart:
\begin{equation} \label{rho}
\rho(H_\text{OBC}) \leq \rho(H_\text{PBC}).
\end{equation}
Note that we are interested in the thermodynamic limit in which all edge-state contribution vanishes. It is evident that Figs.~\ref{fig:amoeba}(a) and (d) satisfy the inequality. The motivation of considering this inequality is a neat inclusion relation in 1D, which states that the OBC spectrum is in the interior of the PBC spectrum \cite{Okuma2020, Zhang2020correspondence}. As the non-Bloch band theory (which the 1D theorem is based on) aligns with the amoeba formulation in 1D, it is natural to ask if a parallel statement can be established in higher dimensions.

We begin the proof with a simple fact about the electrostatic potential. Suppose that a Coulomb potential $\Phi$ is generated by a DOS (charge) distribution $\rho$ in the 2D complex plane, i.e.,
\begin{equation}
    \Phi(E)=\int d^2 E' \, \log |E-E'| \rho(E')    .
\end{equation}
We are interested in the average potential on a circle, say, $|E|=R$, which reads
\begin{equation}
    \bar{\Phi} = \frac{1}{2\pi} \int_0^{2\pi} d\theta \, \Phi \left( R e^{i\theta} \right)  .
\end{equation}
To calculate this average, one just needs to treat all charges inside the circle as distributed uniformly on the circle, and all charges outside as they are. Because of the additivity of the potential, one needs only to prove this property for a single point charge. The contribution of a point charge $q$ with radial coordinate $r<R$ is $\bar{\Phi}=q\log R$, while that of a point charge with $r>R$ is $\bar{\Phi} = q\log r$. This can be readily proved using Jensen's formula Eq.~(\ref{Jensen}). Taking $g(Re^{i\theta}) = R e^{i\theta} - re^{i\alpha}$ in Eq.~(\ref{Jensen}), the left-hand side of the equation is the average potential generated by a point charge at $E'=r e^{i\alpha}$, and Eq.~(\ref{Jensen}) becomes
\begin{equation}
\bar{\Phi}=
\left\{
\begin{aligned}
&\log R, \quad & r<R, &\\
&\log r, \quad  & r>R.
\end{aligned}
\right.\label{Pdef}
\end{equation}

Now we proceed with the spectral radius inequality. From Sec.~\ref{sec:dos}, we know that the Coulomb potential generated by the OBC DOS  is equal to the minimum of the Ronkin function $\Phi_\text{OBC}(E)=\min_{\bm{\mu}} R_{\det(E-h)}(\bm{\mu})$. On the other hand, we can also write the Coulomb potential generated by the PBC DOS in terms of the Ronkin function at $\bm{\mu}=\bm{0}$:
\begin{equation}
    \Phi_{\mathrm{PBC}}(E) = R_{\det(E-h)}(\bm{0}).
\end{equation}
Hence,
\begin{equation}
    \Phi_{\mathrm{OBC}}(E) \leq \Phi_{\mathrm{PBC}}(E).
\end{equation}
It follows that the average satisfies
\begin{equation}
    \bar{\Phi}_{\mathrm{OBC}} \leq \bar{\Phi}_{\mathrm{PBC}}. \label{obcpbc}
\end{equation}
Taking a circle $|E|=R$ that surrounds the whole PBC spectrum, the average potential on this circle is $\bar{\Phi}_{\mathrm{PBC}}=Q\log R$, where $Q$ is the total charge, which is equal to the number of bands of the specific lattice model. Since the total charge of the OBC DOS is also $Q$, the aforementioned electrostatic fact implies that $\bar{\Phi}_{\mathrm{OBC}}\geq Q\log R$, with ``$=$'' reached when the charge density vanishes outside the circle. Combining this with Eq. (\ref{obcpbc}), we obtain the identity $\bar{\Phi}_{\mathrm{OBC}} = Q \log R$. Consequently, the OBC DOS is zero outside the circle, which means that the OBC spectral radius is no larger than the PBC counterpart.

An alternative proof of the spectral inequality is based on Eq. (\ref{rhozero}). We denote the PBC spectral (outer) boundary as $S$, which consists of one or several closed curves (note that the spectrum may also have an inner boundary, which is not included in $S$; for example, an annulus-shaped spectrum has an outer and an inner boundary).  Consider an energy $E$ outside $S$. Since $E$ is not in the PBC spectrum, $\bm{\mu}=\bm{0}$ must belong to the complement of the amoeba of $\det(E-h)$. Furthermore, one can see that $\bm{\mu}=\bm{0}$ locates in the central hole. In fact, in the $|E|\rightarrow\infty$ limit (with $\bm{\mu}=\bm{0}$ fixed), the phase of $\det[E-h(\bbeta)]$ is determined solely by $E$ and independent of $\bm{k}$, and therefore the phase winding number along each $k_j$ circle is zero [cf. Eqs. \eqref{gradient} and \eqref{winding}]. It follows that, for sufficiently large $|E|$, the order $\bm{\nu}=\bm{0}$ at $\bm{\mu}=\bm{0}$, i.e., $\bm{\mu}=\bm{0}$ locates in the central hole. For any $E$ outside $S$, one may connect $E$ to infinity by a path that does not intersect $S$. Since $\bm{\mu}=\bm{0}$ is always in the same amoeba hole as $E$ varies along this path, this amoeba hole must be the central hole. Thus, for any $E$ outside $S$, $\bm{\mu}=\bm{0}$ locates in the central hole of the amoeba. It follows from Eq. \eqref{rhozero} that the OBC DOS $\rho_\text{OBC}(E)=0$ outside $S$. Therefore, the OBC spectrum is enclosed by the PBC spectral boundary. This statement is slightly stronger than the spectral inequality Eq. \eqref{rho}. Particularly, when the energy spectrum is not convex in the complex plane, this statement implies Eq. \eqref{rho} but not vice versa.

We can further extend the above result as follows. Taking $\bm{\mu}$ to be any value other than $\bm{0}$, we readily find that the OBC spectrum is enclosed by any TBC$(\bm{\mu})$ spectrum, where TBC$(\bm{\mu})$ refers to a PBC Hamiltonian generated by symbol $h(e^{\bm{\mu}+i\bm{k}})$, i.e., a Hamiltonian under ``twisted'' periodic boundary condition. We also mention that the development of our theorem is reminiscent of a similar theorem in 1D \cite{Okuma2020},  as they both rely on the winding numbers of the symbol. 

A corollary is immediately implied: The spectral range of the real (or imaginary) part of the OBC bulk spectrum is within its PBC counterpart. To be specific, one has the following general inequalities:
\begin{align}
\max\Re (\epsilon_{n, \text{OBC}}) &\leq \max\Re (\epsilon_{n, \text{PBC}}), \label{Remax} \\
\max\Im (\epsilon_{n, \text{OBC}}) &\leq \max\Im (\epsilon_{n, \text{PBC}}), \label{Immax} \\
\min\Re (\epsilon_{n, \text{OBC}}) &\geq \min\Re (\epsilon_{n, \text{PBC}}), \\
\min\Im (\epsilon_{n, \text{OBC}}) &\geq \min\Im (\epsilon_{n, \text{PBC}}).
\end{align}

In view of the significance of the energy spectrum, the spectral inequalities have various implications, some of which have already been exploited. In one dimension, Eq.~\eqref{Immax} has played an important role in the directional amplification \cite{McDonald2018phase,Xue2021simple,Wanjura2019}. In the context of open quantum systems, the spectral inequalities are closely related to the boundary sensitivity of Liouvillian gap and relaxation time \cite{Song2019,haga2021liouvillian,yang2022liouvillian}. Our general statement and proof of the spectral inequalities lay the groundwork for their higher-dimensional applications.

\section{\label{sec:conc}Concluding remarks}

In this work, we have formulated an amoeba theory of the non-Hermitian skin effect and non-Bloch band theory in arbitrary spatial dimensions. It provides a theoretical framework for studying periodic non-Hermitian systems without the serious dimensional limitation. Among other applications, our theory offers a general yet efficient approach to compute the key physical quantities of non-Hermitian systems, such as the energy spectrum, density of states, and generalized Brillouin zone.

Although the initial version of non-Bloch band theory was formulated under the OBC, the concept of GBZ in 1D is also generalizable to other boundary conditions such as the domain wall cases \cite{Deng2019}. Nevertheless, it seems that the amoeba approach, as we now understand, naturally corresponds to the standard OBC systems. Thus,  we have focused on the OBC case throughout the present paper. Compared to other boundary conditions, the OBC is especially important in many senses. First, the OBC is experimentally the most relevant because realistic systems often have OBC. Second, taking the OBC enables the investigation of both bulk and boundary physics, while other boundary conditions including the PBC are blind to boundary phenomena. Third, even certain measurable physical quantities far from the boundary can be naturally expressed in terms of the GBZ from rather than the conventional BZ associated with PBC \cite{Longhi2019Probing,Longhi2019nonBloch}, though it is in principle free to choose the boundary condition when investigating the physics deep in the bulk. In this sense, the OBC seems to be an advantageous choice even for studying certain bulk physics.

We would also like to remark that not all aspects of this work are mathematically rigorous. Although numerical evidence is supplied whenever a mathematically strict derivation is unavailable, a fully rigorous proof of all our main results is of course desirable. 

In view of the ubiquity of periodic structures in both natural and synthetic systems, it is hoped that our theory can find wide applications in the abundant non-Hermitian phenomena. For example, our formulation can be naturally applied to open quantum systems in the free-particle limit, for which the energy spectrum of the non-Hermitian Liouvillian superoperator determines the dynamics and relaxation \cite{Song2019,haga2021liouvillian}. Our theory immediately enables calculating the relevant quantities beyond 1D. Furthermore, for many-body non-Hermitian systems, our amoeba theory may still be a good starting point for including the interaction effects, which will be left for future work.

\begin{acknowledgments}
This work is supported by NSFC under Grant No. 12125405.
\end{acknowledgments}

\appendix

\section{\label{apd:szego}A brief proof of Szegő's limit theorem}

In this appendix, we sketch a physicist-oriented proof for Szegő's limit theorem Eq.~\eqref{eq:szego}, along with an estimation of the inverse of a Toeplitz matrix. We derive concrete conditions for the theorem to hold and explain their intuitions. This proof is inspired by a parallel treatment for translationally invariant differential operators \cite{Widom1974,Widom1980182}. However, the specificity of Toeplitz matrices on lattice requires additional topological constraints, as stated below.
%


Let us consider a subspace $\Omega$ of the $d$-dimensional Euclidean space. The number of unit cells in $\Omega$ is then $O(L^d)$, and the number of unit cells on the boundary is $O(L^{d-1})$, where $L$ is the linear scale of $\Omega$ (e.g., the side length of a square, or the diameter of a disk).


\begin{figure}[t]
    \centering
    \includegraphics[width=0.8\linewidth]{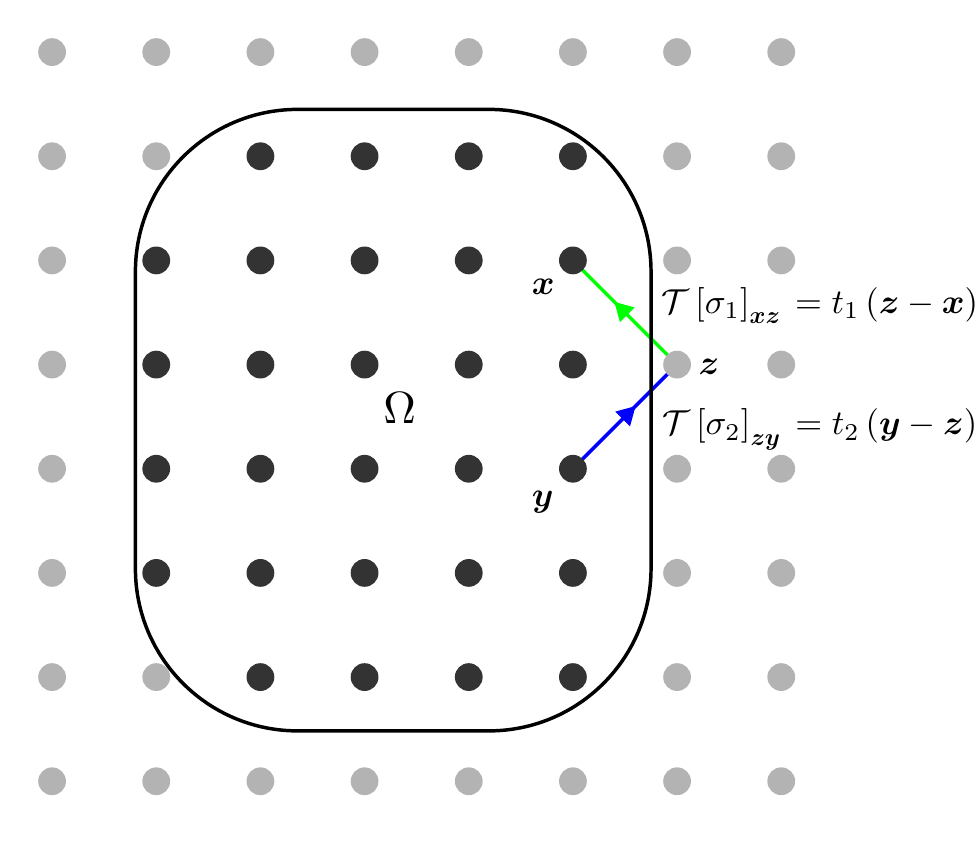}
    \caption{Illustration of Eq.~(\ref{aeq:matele}). The matrix entries of $\mathcal{T}\left[ \sigma_1 \sigma_2 \right] - \mathcal{T}\left[ \sigma_1 \right] \mathcal{T}\left[ \sigma_2 \right]$ are nonzero only when $x$ and $y$ are in a neighborhood of the boundary of $\Omega$. }
    \label{fig:matrixelements}
\end{figure}

Consider two (block-)Toeplitz matrices $\mathcal{T}[ \sigma_1]$ and $\mathcal{T}[ \sigma_2]$ defined in $\Omega$. Both of them are $N\times N$ block matrices, where $N$ is the number of unit cells (labeled by integer-valued coordinates) in $\Omega$. We first establish the following relation:
\begin{equation} \label{aeq:trclass}
    \left\| \mathcal{T}\left[ \sigma_1 \sigma_2 \right] - \mathcal{T}\left[ \sigma_1 \right] \mathcal{T}\left[ \sigma_2 \right] \right\|_{1}
    = O \left( L^{d-1} \right)    ,
\end{equation}
where we use the Schatten norm for operators: $\left\| T \right\|_p$ is the $p$-norm of all singular values $\left\{ s_n \right\}$ of $T$:
\begin{equation}
    \left\| T \right\|_p = \left( \sum_{n} s_n^p \right)^{1/p}  .
\end{equation} For this kind of relation, we use the notation $\mathcal{T}[\sigma_1 \sigma_2] \approx \mathcal{T}[\sigma_1] \mathcal{T}[\sigma_2]$. The $(\bm{x},\bm{y})$ matrix element of $\mathcal{T}[\sigma_1 \sigma_2] - \mathcal{T}[\sigma_1] \mathcal{T}[\sigma_2]$ reads
\begin{equation} \label{aeq:matele}
    \sum_{\bm{z}\notin \Omega}
    t_1 \left( \bm{z}-\bm{x} \right) t_2 \left( \bm{y}-\bm{z} \right)   ,
\end{equation}
where $\bm{x}$, $\bm{y}$ and $\bm{z}$ are integer coordinates; $t_{1,2}$ are the hopping coefficients that appear in $\mathcal{T}[\sigma_{1,2}]$:
\begin{equation}
    t_{1,2}(\bm{n}) = \int_{T^d} \left( \dfrac{d\theta}{2\pi} \right)^d
    \sigma_{1,2}(e^{i\bm{\theta}}) e^{-i\bm{n}\cdot\bm{\theta}}  .
\end{equation}
Note that in Eq.~(\ref{aeq:matele}) $\bm{x},\bm{y}\in \Omega$ while $\bm{z}\in \Omega^c$, where $\Omega^c$ is the complement of $\Omega$ in the Euclidean space. The terms $\sum_{\bm{z}\in \Omega}t_1  ( \bm{z}-\bm{x}) t_2  ( \bm{y}-\bm{z})$ appear in both $\mathcal{T}[\sigma_1 \sigma_2]$ and $\mathcal{T}[\sigma_1] \mathcal{T}[\sigma_2]$ and therefore do not contribute to their difference.



It is clear that each term in Eq.~(\ref{aeq:matele}) can be nonzero only when both $\bm{x}$ and $\bm{y}$ are near the boundary (see also Fig.~\ref{fig:matrixelements}). Thus, it is intuitive that the matrix norm of $\mathcal{T}[\sigma_1 \sigma_2] - \mathcal{T}[\sigma_1] \mathcal{T}[\sigma_2]$ is of order $O(L^{d-1})$. A more formal proof is based on Hölder's inequality for matrix norms, which states that $\left\| AB \right\|_1\leq \left\| A \right\|_p \left\| B \right\|_q$ holds for all $p\in[1,\infty]$ and $q$ satisfying $1/p+1/q=1$. Defining $\left( T_1 \right)_{\bm{x}, \bm{z}} = t_1(\bm{z}-\bm{x})$ that maps $\Omega^c$ to $\Omega$ and $\left( T_2 \right)_{\bm{z}, \bm{y}} = t_2(\bm{y}-\bm{z})$ that maps $\Omega$ to $\Omega^c$, we see that Eq.~(\ref{aeq:matele}) becomes $(T_1T_2)_{\bm{x},\bm{y}}$. According to Hölder's inequality, we have
\begin{equation}
    \left\| T_1 T_2 \right\|_1
    \leq \left\| T_1 \right\|_2 \left\| T_2 \right\|_2,
\end{equation}
and we are left only with showing $\left\| T_{j} \right\|_2^2 = O(L^{d-1})$ for $j=1,2$. Recalling that the singular values of $T_1$ are the eigenvalues of $\sqrt{T_1 T_1^\dagger}$, $\left\| T_1 \right\|_2^2$ by definition equals the trace of $T_1 T_1^\dagger$, whose elements read
\begin{equation}
    \left( T_1 T_1^\dagger \right)_{\bm{x}, \bm{x'}} = \sum_{\bm{z}\in\Omega^c} t_1(\bm{z}-\bm{x}) t_1^\dagger(\bm{z}-\bm{x'})    .
\end{equation}
Hence
\begin{align}   \label{aeq:derivet1}
    \left\| T_1 \right\|_2^2
    &= \tr T_1 T_1^\dagger  \notag\\
    &= \sum_{\bm{x}\in\Omega,\bm{z}\in\Omega^c} \tr t_1(\bm{z}-\bm{x}) t_1^\dagger (\bm{z}-\bm{x})  \notag\\
    &= \sum_{\bm{x}\in\Omega,\bm{z}\in\Omega^c} \left\| t_1(\bm{z}-\bm{x}) \right\|_2^2 \notag\\
    &= \sum_{\bm{u}} \left\| t_1(\bm{u}) \right\|_2^2 \operatorname{vol} \left( (\Omega+\bm{u}) \cap \Omega^c \right)   .
\end{align}
In the last line we substitute $\bm{z}-\bm{x}$ by $\bm{u}$. The volume accounts for the multiplicity of the term $\left\| t_1(\bm{u}) \right\|_2^2$, which is the number of lattice points in the overlap of $\Omega$ translated by $\bm{u}$ and the complement $\Omega^c$. It is roughly $|\bm{u}|$ times the surface area of $\Omega$ denoted by $|\partial\Omega|$.  For short-ranged Toeplitz matrices, we have $t_j(\bm{u})=0$ when $|\bm{u}|$ is beyond the hopping range. Therefore, only a boundary layer contributes to the summation over $\bm{u}$, which means
\begin{equation}
    \left\| T_1 \right\|_2^2
    = O \left( L^{d-1} \right)    .
\end{equation}
A similar estimation holds for $T_2$. These two estimates together prove Eq.~(\ref{aeq:trclass}). Notably, Eq.~\eqref{aeq:trclass} holds even if we add disorders (which is considered in Sec. \ref{sec:dos}) on the boundary of $\mathcal{T}\left[ \sigma_1 \right]$ and $\mathcal{T}\left[ \sigma_2 \right]$, since the difference on the left-hand side is still localized near the boundary. 

A caveat arises when either $\mathcal{T}\left[ \sigma_1 \right]$ or $\mathcal{T}\left[ \sigma_2 \right]$ is not short ranged, meaning, e.g., $t_1(\bm{u})$ is nonzero even for large $|\bm{u}|$. For a large system size, Eq.~(\ref{aeq:derivet1}) is less than
\begin{equation}
    \sum_{\bm{u}} \left\| t_1(\bm{u}) \right\|_2^2 \left|\bm{u}\right| \left| \partial\Omega \right|    .
\end{equation}
Thus, a necessary condition for Eq.~(\ref{aeq:trclass}) and all the following theorems to hold is that
\begin{equation}    \label{aeq:shortrange}
    \sum_{\bm{u}} \left|\bm{u}\right| \left\| t(\bm{u}) \right\|_2^2 < \infty
\end{equation}
is satisfied for all Toeplitz matrices in consideration. This tells us that our theorems also hold for certain long-ranged Hamiltonians, as long as the hopping decays sufficiently fast at long distance.

Now let $\sigma_1=\sigma$, $\sigma_2=\sigma^{-1}$ and assume that $\sigma$ is invertible, we have
\begin{equation}
    \left\| I - \mathcal{T}\left[ \sigma \right] \mathcal{T}\left[ \sigma^{-1} \right] \right\|_{1}
    = O \left( L^{d-1} \right)    ,  \label{aeq:inver}
\end{equation}
where $I$ is the identity matrix. Intuitively, multiplying $\mathcal{T}[\sigma]^{-1}$ on the left, we would expect the following asymptotic inversion formula:
\begin{equation} \label{aeq:inv}
    \left\| \mathcal{T} \left[ \sigma \right]^{-1} - \mathcal{T}\left[ \sigma^{-1} \right] \right\|_{1}
    = O \left( L^{d-1} \right)    .
\end{equation}
Nevertheless, this multiplication should not be done without discretion. To be precise, we use Hölder's inequality again,
\begin{align}
    &\left\| \mathcal{T} \left[ \sigma \right]^{-1} - \mathcal{T}\left[ \sigma^{-1} \right] \right\|_{1}   \notag \\
    &\leq
    \left\| \mathcal{T}[\sigma]^{-1} \right\|_\infty \left\| I - \mathcal{T}\left[ \sigma \right] \mathcal{T}\left[ \sigma^{-1} \right] \right\|_{1}    .
\end{align}
Therefore, a condition for Eq.~(\ref{aeq:inv}) is that $\left\| \mathcal{T}[\sigma]^{-1} \right\|_\infty$, which is the largest of its singular values, is finite in the thermodynamic limit $L\rightarrow\infty$. This amounts to asking if $\mathcal{T}[\sigma]$ itself has a zero singular value when $L\rightarrow\infty$. Since most of its singular values form bulk bands and have a finite gap from zero, the zero singular values should be of certain topological origin. They are intimately related to the index theorem, or the bulk-boundary correspondence in free systems. If the symbol has a nontrivial topological index, then the corresponding Toeplitz matrix has either nonempty kernel or nonempty cokernel. Either case implies that $\mathcal{T}[\sigma] \mathcal{T}[\sigma]^\dagger$ is not fully ranked and has a zero eigenvalue. As to 1D Toeplitz matrices, the topological index is well known to be the negative of the winding number of $\det\sigma(e^{i\theta})$ \cite{douglas1998banach}. To ensure that the Toeplitz matrix does not have zero singular values, we impose a condition that the symbol $\sigma$ is homotopically trivial \footnote{As a heuristic example, consider $\sigma(e^{i \theta})=\epsilon + e^{i\theta}$. For $\epsilon=10$, the symbol is homotopic to a constant symbol and the winding number is zero. One can check that both Eq. (\ref{aeq:inver}) and Eq. (\ref{aeq:inv}) hold. As a comparison, the $\epsilon=1/10$ case has winding number $1$, meaning that the symbol is not homotopic to a constant symbol. Accordingly, Eq.~(\ref{aeq:inver}) is satisfied, while Eq.~(\ref{aeq:inv}) is not. Instead, it is homotopic to a unitary translation operator, whose symbol is $\sigma(e^{i \theta}) = e^{i\theta}$. It is now obvious that a unitary translation maps the leftmost mode to zero, thus having one zero singular value. Being in the same topological sector, the matrix with $\epsilon=1/10$ also has a zero singular value (with a correction exponentially small in size). }. This means that there exists a continuous path in the space of invertible symbols subject to Eq.~(\ref{aeq:shortrange}), connecting the symbol $\sigma(e^{i\bm{\theta}})$ to the constant symbol $\sigma_{\text{const}}(e^{i\bm{\theta}}) = 1$. This requirement may be too strong for now, but we see in a moment that it arises again when defining the matrix logarithm.

The asymptotic inversion formula Eq.~(\ref{aeq:inv}) aims at finding a good surrogate for the inverse of $\mathcal{T}[\sigma]$ in the bulk, $\mathcal{T}[\sigma^{-1}]$, which is calculable by integration alone.
We remark that Widom also proposed an explicit expression for the subleading term in $\mathcal{T}[\sigma]^{-1} = \mathcal{T}[\sigma^{-1}] + O(L^{d-1}) + o(L^{d-1})$ in one-dimensional case \cite{Widom1974, WIDOM19761}, and also in higher-dimensional continuous case \cite{Widom1980182}. It is not used in the current paper and therefore omitted. Szegő's limit theorem also naturally inherits an expression for the subleading term $O(L^{d-1})$.


Next, we move on to Szegő's limit theorem Eq.~\eqref{eq:szego}. Since $\log\det A = \tr\log A$, it is equivalent to the following trace formula:
\begin{equation}
    \tr \log \left( \mathcal{T} \left[ \sigma \right] \right)
    = N \int_{T^d} \left( \dfrac{d\theta}{2\pi} \right)^d \tr \log \left( \sigma \left( e^{i\bm{\theta}} \right) \right)
    + O \left( L^{d-1} \right)    ,
\end{equation}
where $N$ is the number of lattice points in $\Omega$. In order to properly define a logarithm for the symbol, we must rely on the homotopy path imposed earlier. Let $\sigma_{\lambda}(e^{i\bm{\theta}})$ be one of these paths, which is a continuous function from $\lambda\in [0,1]$ to the space of invertible symbols. To ensure the path independence of the integral in Eq.~\eqref{aeq:logdef} below, we further require that $[\sigma'_{\lambda}, \sigma_\lambda^{-1}] = 0$. 
For example, a possible construction of such a homotopy path in the context of taking $\sigma(e^{i\bm{\theta}}) = E - h(e^{\bm{\mu}+i\bm{\theta}})$ is as follows. We can find a continuous path $E_\lambda$ connecting $E_0 = E$ and $E_1 = \infty$ that avoids the spectral range of $h(e^{\bm{\mu}+i\bm{\theta}})$ (with $E_\lambda\neq E-1$), then the path of symbol can be taken as
\begin{equation}
    \sigma_{\lambda}(e^{i\bm{\theta}}) = \frac{ E_\lambda - h(e^{\bm{\mu}+i\bm{\theta}}) }{ E_\lambda - E + 1 } .
\end{equation}
A consequence of trivial homotopy is that the winding number of $\sigma$ along any circle on $T^d$ is identically zero.

Moreover, it is an interesting question what role is played by point-gap topological invariants other than the winding numbers Eq.~\eqref{winding}. Our current conjecture is that the (1D) winding numbers are all that are related to the appearance of NHSE. Mathematically, it means that the Szegő's limit theorem holds even if other topological numbers are nonzero. For example, the 3D winding number can protect topological surface states in non-Hermitian band systems, but it does not induce NHSE \cite{Denner2021}. Possibly the subleading terms in Szegő's limit theorem will be helpful in quantifying such surface states. 

The logarithm in Szegő's limit theorem is now defined as
\begin{equation}    \label{aeq:logdef}
    \log\sigma = \int_0^1 \sigma_\lambda' \sigma_\lambda^{-1} \,d\lambda    ,
\end{equation}
\begin{equation}
    \log\mathcal{T}\left[ \sigma \right] = \int_0^1 \mathcal{T} \left[ \sigma_\lambda' \right] \mathcal{T} \left[ \sigma_\lambda \right]^{-1} \,d\lambda    ,
\end{equation}
where a prime denotes a derivative with respect to $\lambda$. Using Eqs.~\eqref{aeq:inv} and \eqref{aeq:trclass}, we deduce
\begin{align}
    \log \mathcal{T} \left[ \sigma \right]
    &\approx \int_0^1 \mathcal{T} \left[ \sigma_\lambda' \right] \mathcal{T} \left[ \sigma_\lambda^{-1} \right] \,d\lambda  \notag\\
    &\approx \int_0^1 \mathcal{T} \left[ \sigma_\lambda' \sigma_\lambda^{-1} \right] \,d\lambda  \notag\\
    &= \mathcal{T} \left[ \log\sigma \right]    .
\end{align}
The reason why we focus on the $1$-norm becomes clear now, since $\tr A \leq \left\| A \right\|_1$ for any matrix or operator $A$. Taking the trace on both sides, Szegő's limit theorem is proved, because the trace of the matrix on the last line is by definition
\begin{align}
    \tr \mathcal{T} \left[ \log\sigma \right]
    &= \sum_{x\in\Omega} \int_{T^d} \left( \dfrac{d\theta}{2\pi} \right)^d \tr \log \left( \sigma \left( e^{i\bm{\theta}} \right) \right)   \notag\\
    &= N \int_{T^d} \left( \dfrac{d\theta}{2\pi} \right)^d \tr \log \left( \sigma \left( e^{i\bm{\theta}} \right) \right)   .
\end{align}

\section{\label{apd:basis}Invariance under change of basis}

An advantage of our method exploiting the amoeba and the Ronkin function is that they are invariant under linear transformations on the basis of the reciprocal space. This fact guarantees that all physical outcomes are independent of a specific choice of basis in the Brillouin zone. We show this invariance in this appendix.

We examine the effect of changing the basis in the reciprocal lattice. Let a basis of the reciprocal lattice be $\{ \bm{b}_1, \dots, \bm{b}_d \}$. An equivalent basis $\{ \bm{b}'_j \}$ is then related to $\{ \bm{b}_j \}$ by
\begin{equation}
    \begin{pmatrix}
        \bm{b}_1 & \cdots & \bm{b}_d
    \end{pmatrix}
    =
    \begin{pmatrix}
        \bm{b}_1' & \cdots & \bm{b}_d'
    \end{pmatrix}
    U.
\end{equation}
To ensure that the new basis spans the entire lattice, we require all elements of $U_{d\times d}$ to be integers, and $\det U = \pm 1$. Thus, $U$ is an element drawn from $\mathrm{GL}(d, \mathbb{Z})$.

Let the real-space Hamiltonian be
\begin{equation}
    H = \sum_{\bm{x}, \bm{y}} \ket{\bm{x}} t_{\bm{y}-\bm{x}} \bra{\bm{y}}    ,
\end{equation}
where we suppress the intracell indices for notational simplicity, but they can be restored without effort. The Bloch Hamiltonian (also known as the symbol) then reads
\begin{equation}
    h(\beta_1, \dots, \beta_d)
    = \sum_{\bm{n}} t_{\bm{n}} \beta_1^{\bm{n}\cdot\bm{b}_1} \dots \beta_d^{\bm{n}\cdot\bm{b}_d}    .
\end{equation}
We may write it alternatively as a function of the complex momentum $\tilde{k}_j=-i\log\beta_j$:
\begin{equation}
    h(e^{i\tilde{k}_1}, \dots, e^{i\tilde{k}_d})
     = \sum_{\bm{n}} t_{\bm{n}} \exp i \left( \tilde{k}_1\bm{n}\cdot\bm{b}_1 + \dots + \tilde{k}_d\bm{n}\cdot\bm{b}_d \right)    .
\end{equation}

After a $U$ transformation, the new Bloch Hamiltonian becomes
\begin{equation}
    h'(e^{i\tilde{k}'_1}, \dots, e^{i\tilde{k}'_d})
     = \sum_{\bm{n}} t_{\bm{n}} \exp i \left( \tilde{k}'_1\bm{n}\cdot\bm{b}'_1 + \dots + \tilde{k}'_d\bm{n}\cdot\bm{b}'_d \right)    .
\end{equation}
We immediately see that $h(e^{i\tilde{k}_1}, \dots, e^{i\tilde{k}_d}) = h'(e^{i\tilde{k}'_1}, \dots, e^{i\tilde{k}'_d})$ as long as $\tilde{k}_1\bm{b}_1 + \dots \tilde{k}_d\bm{b}_d = \tilde{k}'_1\bm{b}'_1 + \dots \tilde{k}'_d\bm{b}'_d$. In other words, $h$ is an invariant scalar function. It follows that $h$ is invariant if the wave vector transforms as
\begin{equation}
    \begin{pmatrix}
        \tilde{k}_1' \\ \vdots \\ \tilde{k}_d'
    \end{pmatrix}
    =U
    \begin{pmatrix}
        \tilde{k}_1 \\ \vdots \\ \tilde{k}_d
    \end{pmatrix} .
\end{equation}

Therefore, the amoeba (coordinates being $\mu_j = -\Im \tilde{k}_j$) undergoes the same linear transformation $U$ when we change the basis. Its properties of physical importance, especially the existence of the central hole, are hence unchanged under a change of basis. Furthermore, for the Ronkin function the following identity holds:
\begin{equation}
    R_{ \det(E-h) } (\bm{\mu}) = R_{ \det(E-h') } (U\bm{\mu})    ,
\end{equation}
where $\bm{\mu}$ stands for the column vector $(\mu_1,\dots,\mu_d)^T$. This is because, by definition,
\begin{equation}
    \begin{aligned}
        &R_{ \det(E-h') } (U\bm{\mu}) \\
        &= \int_{T^d} \left( \frac{d\theta}{2\pi} \right)^d \log \left| \det \left( E-h'(e^{U\bm{\mu}+i\bm{\theta}}) \right) \right|    .
    \end{aligned}
\end{equation}
Renaming $\bm{\theta}\to U\bm{\theta}$ and noting that the integration measure is unchanged, we find the right-hand side to be equal to $R_{ \det(E-h) } (\bm{\mu})$.

\bibliography{dirac,references}

\end{document}